\documentclass[12pt,preprint]{aastex}
\begin{document}

\shorttitle{X-rays from W40}
\shortauthors{Kuhn et al.}

\slugcomment{Accepted for publication in the Astrophysical Journal 10/24/2010}

\title{A Chandra Observation of the Obscured Star-Forming Complex W40}

\author{Michael A.\ Kuhn\altaffilmark{1}, Konstantin V.\ Getman\altaffilmark{1}, Eric D.\
Feigelson\altaffilmark{1,2}, Bo\ Reipurth\altaffilmark{3}, Steven A.\ Rodney\altaffilmark{4}, 
Gordon P.\ Garmire\altaffilmark{1}}

\altaffiltext{1}{Department of Astronomy \& Astrophysics, 525
Davey Laboratory, Pennsylvania State University, University Park
PA 16802} \altaffiltext{2}{Center for Exoplanets and Habitable Worlds, 525
Davey Laboratory, Pennsylvania State University, University Park
PA 16802} \altaffiltext{3}{University of Hawaii, 
Institute for Astronomy, 640 N. Aohoku Pl., Hilo, HI 96720}
\altaffiltext{4}{Johns Hopkins University, 3001 San Martin Drive, Baltimore, MD 21218}

\email{mkuhn1@astro.psu.edu, gkosta@astro.psu.edu}

\begin{abstract}

The young stellar cluster illuminating the W40 H II region, one of the nearest massive star forming regions, has been observed with the ACIS detector on board the Chandra X-ray Observatory. Due to its high obscuration, this is a poorly-studied stellar cluster with only a handful of bright stars visible in the optical band, including three OB stars identified as primary excitation sources. We detect 225 X-ray sources, of which 85\% are confidently identified as young stellar members of the region. Two potential distances of the cluster, 260~pc and 600~pc, are used in the paper. Supposing the X-ray luminosity function to be universal, it supports a 600~pc distance as a lower limit for W40 and a total population of at least 600 stars down to 0.1~M$_{\odot}$ under the assumption of a coeval population with a uniform obscuration. In fact, there is strong spatial variation in $K_s$-band-excess disk fraction and non-uniform obscuration due to a dust lane that is identified in absorption in optical, infrared and X-ray. The dust lane is likely part of a ring of material which includes the molecular core within W40. In contrast to the likely ongoing star formation in the dust lane, the molecular core is inactive. The star cluster has a spherical morphology, an isothermal sphere density profile, and mass segregation down to 1.5~M$_{\odot}$. However, other cluster properties, including a $\la 1$~Myr age estimate and ongoing star formation, indicate that the cluster is not dynamically relaxed. X-ray diffuse emission and a powerful flare from a young stellar object are also reported. 



\end{abstract}


\section{Introduction \label{introduction_section}}

\begin{figure}
\centering
\includegraphics[angle=0.,width=7.0in]{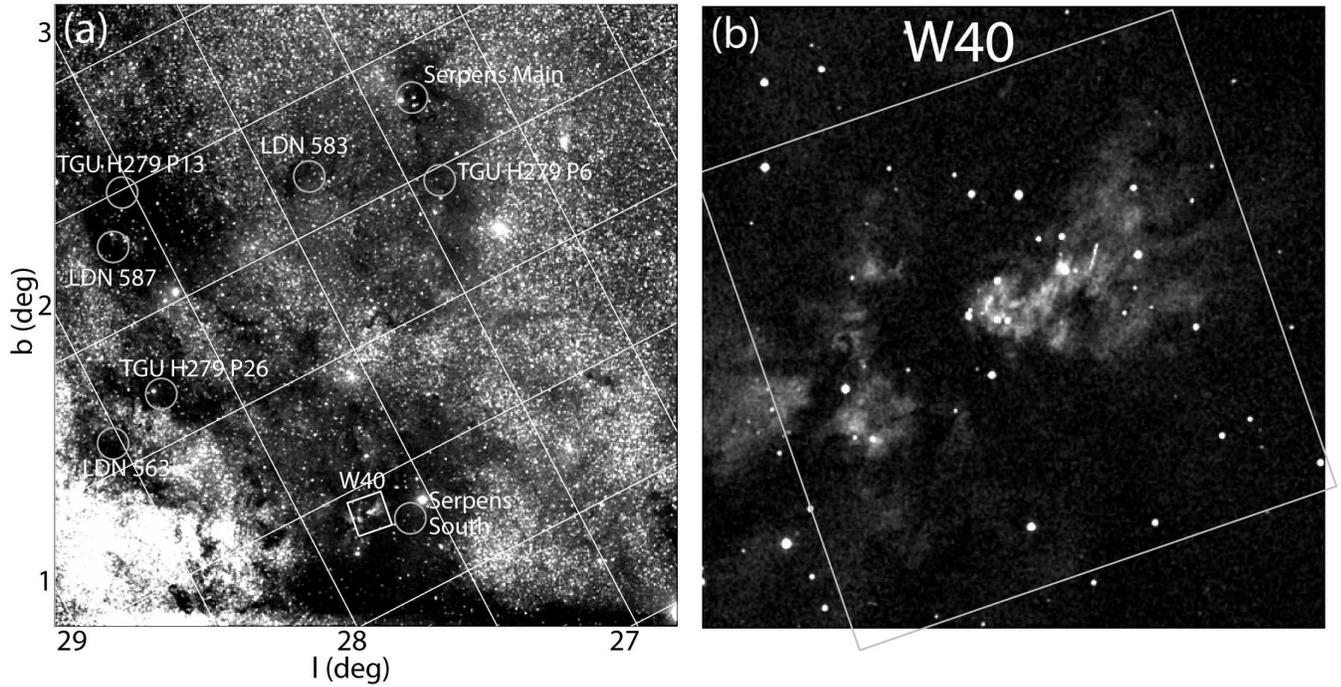}
\caption{(a) 5$\,^{\circ}\times$5$\,^{\circ}$ DSS Red image of the neighborhood of W40, including Serpens South and Serpens Main. Nearby dark clouds and star forming regions are indicated by circles. The {\it Chandra} ACIS-I field of view is indicated by a white square. (b) Enlarged image of the ACIS-I field of view. Nebulous optical emission and a few stars projected on W40 may be seen.\label{fig_dss}}
\end{figure}

The W40 \citep{Westerhout58} complex is a star-formation region in the constellation Serpens. The complex is made up of the molecular cloud G28.8+3.5 \citep{Goss70} with a diameter of $\sim1\,^{\circ}$, the dense molecular core DoH 279-P7 \citep{Dobashi05} with a diameter of $\sim20^{\prime}$, an OB star cluster, and an H II blister region. The complex is behind a screen of $A_V\ge5$~mag of extinction, so only a handful of stars can be seen in the optical band (Figure \ref{fig_dss}b). Because of the high absorption, the stellar cluster has been little studied. The cluster is relatively nearby, but the distance has not been sufficiently constrained and estimates vary from 300~pc to 900~pc. Previous observations show that W40 is partially embedded in its natal molecular cloud \citep[e.g.][]{Vallee87}, which indicates that the cluster is no more than a few million years old.

Infrared (IR) has been the primary wavelength for studies of star formation. However, certain abilities of X-ray studies make observations particularly useful for examining young stellar clusters (YSCs). X-rays effectively distinguish cluster members from foreground or background stars since young stellar objects have typical X-ray luminosities 10$^3$-10$^5$ times greater than main-sequence stars \citep[e.g.,][]{Preibisch05b}. The high penetrating power of hard X-rays is required to study young stars which are often embedded in natal molecular material. The {\it Chandra} Orion Ultradeep Project \citep[COUP;][]{Getman05a} reveals stars with up to $500$~mag of visual-band extinction \citep{Grosso05}. Another advantage of studies in the X-ray is that bright nebulosity, which may be a source of confusion in the IR or optical, is avoided at X-ray wavelengths, yielding greater source detection sensitivity. Differences between X-ray brightness of T-Tauri and OB stars are not so dramatic as in optical and in IR bands, so X-rays often allow detection of low mass pre-main sequence (PMS) stars close to OB stars. X-ray observations also preferentially detect pre-main-sequence (PMS) stars without disks (Class III), unlike the IR, where stars possessing circumstellar disks (Class II or I/0) are preferentially detected due to IR excesses. Thus, X-ray observations reduce biases present in IR surveys of young stellar objects (YSOs), and X-ray and IR surveys are complementary to each other. 

The ability to observe stars with large optical extinctions is essential in any study of W40 due to the large amount of obscuring material along the line of sight. The high spatial resolution of {\it Chandra} is also necessary to observe stars in the center of the cluster. The X-ray is used to select cluster members for our analysis of W40, and stars with masses down to 0.1~M$_{\odot}$ are detected with the data from the 38~ksec {\it Chandra} observation. The analysis of sources of contamination for X-ray observations of stellar clusters and effective methods to pick out cluster stars from samples that include foreground stars and extragalactic objects are described by \citet{Getman06}. Once a list of W40 cluster members has been created, both X-ray and other wavelengths are used to determine properties of stars in the sample. In addition, the spatial distribution of the cluster stars is analyzed. Other interesting phenomena are present in the X-ray data, including diffuse emission and a large flare from an intermediate mass young star.

The observation and source lists are described in \S \ref{section_obs}. Optical, infrared, and radio counterparts are identified in \S \ref{oir_section} and near-IR (NIR) stellar properties are derived. Stellar and extragalactic contaminants in our source lists are identified in \S \ref{membership_section}. X-ray properties are derived in \S \ref{spectral_section}. Confirmatory results relating X-ray and IR properties are presented in \S \ref{confirmation_section}. The X-ray luminosity function (XLF), cluster distance, total cluster population, intrinsic disk fraction, and age are discussed in \S \ref{section_XLF}. The initial mass function (IMF) is given in \S \ref{imf_section}. Massive star candidates are discussed in \S \ref{section_OB}. Spatial structure is discussed in \S \ref{spatial_section} in which the detection of an absorbing filament, spatial segregation of high versus low-mass stars, and spatial inhomogeneity in disk fraction of stellar populations is reported. In \S \ref{interesting_section} we end with the analysis of X-ray diffuse emission and mention a powerful flare from an intermediate mass young star.

\subsection{Past Studies of W40 \label{past_section}}


The distance to W40 has been controversial. Radial velocity based distance measurements \citep{Reifenstein70,Downes70,Radhakrishnan72} using the \citet{Schmidt65} model of the galaxy collectively suggest a distance of $600\pm300$~pc \citep{Vallee87}. \citet{Shaver70} also find a distance of 700~pc using galactic kinematics. Other distance measurements fall into this range too. \citet{Crutcher82} find a distance of 400~pc assuming a spectral type for the brightest star of B0V. \citet{Smith85} find a distance of 700~pc using measurements of W40's stellar component in radio/infrared continuum. \citet{Kolesnik83} use OH absorption-line measurements of the cloud and calculate a distance of 600~pc. These estimates range from 400~pc to 700~pc, and we adopt a distance of $\sim$600~pc in line with \citet{Rodney08} and investigate the validity of this distance in the analysis of the {\it Chandra} data. W40 has Galactic coordinates $l=28.8\,^{\circ}$, $b=3.5\,^{\circ}$. The fairly high Galactic latitude would place the cluster more than 50~pc above the Galactic plane if the distance were greater than 800~pc. 

The Aquila Rift, which contains the young star cluster Serpens South \citep{Gutermuth08}, is projected near W40 in the sky (Serpens South is 0.5$^{\circ}$ to the west; see Figure \ref{fig_dss}a) and is only 260$\pm$37~pc distant \citep{Straizys96}, which is much closer than typical distance estimates for W40. However, W40 and Serpens South are usually regarded as separate objects due to their different distances \citep{Straizys03,Gutermuth08}, and 600$-$700~pc is the most commonly cited distance for W40 \citep[cf.][]{Zhu06,Zinnecker07,Rodriguez_in_prep}. The nearby Serpens Main cluster is 3$^{\circ}$ north of W40 and is at a third distance of 415$\pm$15~pc \citep{Dzib10}, established through parallax measurements of its brightest star. \citet{Dzib10} notes that the superposition of the three distinct star-forming regions is not unexpected since the line of sight passes just above the Galactic plane. Alternatively, \citet{Bontemps10} assume that W40, the Aquila Rift, and Serpens Main are parts of the same star-forming region and are therefore at the distance of 260~pc for the Aquila Rift. This paper will comment on implications of this lower distance for W40.


The core of the W40 molecular cloud was mapped by \citet{Zeilik78} in $^{12}$CO and $^{13}$CO emissions. The mass of the core is estimated to be $\sim 200-300$~M$_{\odot}$ by \citet{Zhu06}, and \citet{Rodney08} report a probable mass for the entire cloud of $\sim10^4$~M$_{\odot}$. The center of the star cluster is located at the east edge of the molecular core. Three probable OB stars are identified by \citet{Smith85} as important ionizing sources for the cluster. There is evidence for circumstellar material around two of these stars detected by \citet{Smith85} and \citet{Vallee94}. An observation of W40 by the {\it Midcourse Space Experiment} (MSX) reveals the H II region to have an hourglass shape, oriented with one lobe to the southeast and another to the northwest,  $17^{\prime}\times30^{\prime}$ in extent \citep[see Figure 6 in][]{Rodney08}. The cluster core is located just to the northwest of the waist of the hourglass. \citet{Bontemps10}, who assume that W40 and Serpens South are part of the same star-forming complex at a distance of 260~pc, report $\sim$200 protostellar candidates in {\it Herschel} observations. A weak outflow is detected in CO lines by \citet{Zhu06} centered on the molecular core, although the driving source is unseen.

\section{Chandra Observation and Source List \label{section_obs}}
\subsection{Observation and Data Reduction}
The {\it Chandra} X-ray Observatory (CXO) observed W40 for $\sim$38~ks during August, 2007 using the ACIS camera \citep{Garmire03}. The observation has a nominal pointing of R.A.$=$ 18$^{\rm h}$ 31$^{\rm m}$ 23.\hspace{-0.6ex}$^{\rm s}$8 and decl.$=$ -2$^{\circ}$ 05$^{\prime}$ 59$^{\prime\prime}$ (J2000), the approximate center of the W40 cluster, and a roll angle of 251 degrees. Data from the imaging array (ACIS-I) are analyzed. The ACIS-S3 chip was also operational, but these data are omitted since the point spread function is degraded due to the large off-axis angle. The data were taken in Very Faint telemetry mode to improve sensitivity to faint sources. No variations in background levels are detected. Properties of the telescope and detector are available via the Proposers' Observatory Guide \citep{cxc09}.

A data reduction protocol that has been used for {\it Chandra} ACIS-I observations of other young-star clusters is followed (see Appendix B of \citet{Townsley03} for more details on the steps used for data reduction). Briefly, using the tool {\it acis\_process\_events} from the CIAO3.4 software package \citep{CIAO}, the latest calibration information on time-dependent gain and a Penn State bad pixel mask are applied, background event candidates are identified in the very faint mode, and the data are corrected for CCD charge transfer inefficiency. Using the {\it acis\_detect\_afterglow} tool, additional afterglow events not detected with the standard Chandra X-ray Center (CXC) pipeline are flagged. The event list is cleaned with the ``grade'' (only ASCA grades 0,2,3,4,6 are accepted), ``status''\footnote{As some of the background and afterglow event candidates from bright sources are real they are retained for our spectral analysis, and the non-controversial status=0 filter is applied only to build the event list for source search.}, ``good-time interval'', and energy filters. A small astrometric correction is applied based on matches between the eight brightest {\it Chandra} sources with small off-axis angles and optical/infrared sources from the Naval Observatory Merged Astrometric Dataset \citep[NOMAD;][]{Zacharias04a}. The slight point-spread function (PSF) broadening from the CXC software position randomizations was removed. 

\subsection{Source Detection and Characterization \label{phot_section}}

Figure \ref{fig_cluster}c,d shows the ACIS-I field in which more than 200 X-ray sources are seen. Candidate X-ray sources in the {\it Chandra} data are detected using an aggressive method that minimizes missed sources but allows spurious sources. Data images and exposure maps with spatial resolutions 0.5$^{\prime\prime}$ and 1$^{\prime\prime}$ pixel$^{-1}$ are used for source detection on the center of ACIS-I and the full ACIS-I, respectively. A wavelet-based source detection procedure \citep[$wavdetect$,][]{Freeman02} is run with a threshold of P=10$^{-5}$. Sources are detected on a soft (0.5-2.0 keV), a hard (2.0-8.0 keV), and the total (0.5-8.0 keV) band images. The images are examined visually to locate other candidate sources, mainly close doubles and sources near the detection threshold. A total-band image reconstruction is generated using the maximum likelihood method to search for additional weak sources in the center region \citep{Broos10}. 

Refined source positions and the final source list are established using the \anchor{http://www.astro.psu.edu/xray/docs/TARA/ae_users_guide.html}{{\em ACIS Extract}} software package\footnote{The {\em ACIS Extract} software package and User's Guide are available online at \url{http://www.astro.psu.edu/xray/docs/TARA/ae\_users\_guide.html}.} \citep{Broos02,Broos10}. Photons are extracted within polygonal contours of $\sim 90$\% ($\sim 70$\% for crowded sources) encircled energy using position-dependent models of the point-spread function (PSF). Background extraction is performed locally and takes into account spatial variations due to PSF wings of nearby point sources \citep{Broos10}. Production of the final source list follows the simple procedure of \citet{Getman06,Getman07,Getman08a}: the list of candidate sources is trimmed to omit sources with fewer than $\sim4$ estimated source background-subtracted counts ($NC_t/{\rm PSFfrac}\la 4$; columns 7 and 11 in Table \ref{tbl_phot}). The final catalog has 225 sources.

The {\em ACIS Extract} package is also used to construct source and background spectra, compute redistribution matrix files (RMFs) and auxiliary response files (ARFs), construct light curves and time-energy diagrams, perform a Kolmogorov-Smirnov (K-S) variability test, compute photometric properties, and perform automated spectral grouping and fitting. The following inferred source photometric properties are presented in Table \ref{tbl_phot}: source coordinates, off-axis angle, net and background counts within the extraction region, source significance assuming Poisson statistics, effective exposure (correcting for telescope vignetting, satellite dithering, and bad CCD pixels), median energy after background subtraction, a variability indicator extracted from the one-sided K-S statistic, and occasional anomalies related to positions on chip gaps or field edges. Net counts has a range 4$-$1000 and typical values of $\sim$ 20. Median energy has a range of 1$-$5~keV and typical values around 2.4~keV.

\section{Optical, Infrared, and Radio Counterparts \label{oir_section}}
\subsection{Identifications \label{identifications_section}}

\begin{figure}
\centering
\includegraphics[angle=0.,width=7.0in]{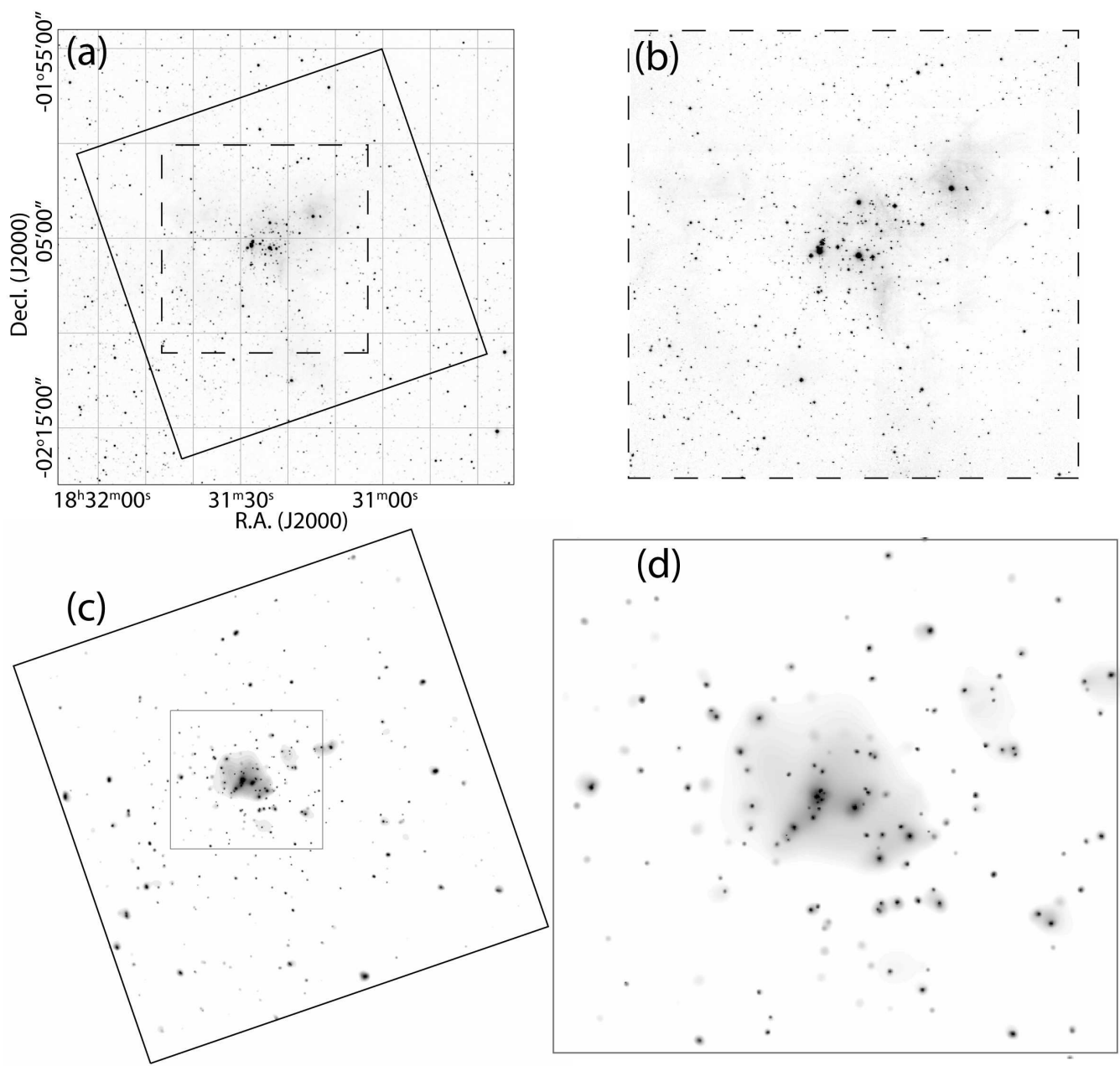}
\caption{(a) $24^{\prime}\times24^{\prime}$ 2MASS $K_s$-band and (b) $11^{\prime}\times11^{\prime}$ UKIRT $K$-band images of the W40 cluster. The solid line indicates the {\it Chandra} ACIS-I field of view and the dashed line indicates the UKIRT field of view. (c) Adaptively smoothed $17^{\prime}\times17^{\prime}$ ACIS-I image of the W40 cluster. Smoothing has been performed on the total 0.5-8.0 keV band at the 2.5 $\sigma$ level, and the grey scale is logarithmic. Most of the 225 X-ray sources are visible. (d) The $6^{\prime}\times5.6^{\prime}$ center of the previous image is shown in greater detail.\label{fig_cluster}}
\end{figure}

{\it Chandra} source positions were compared with source positions from the 2-Micron All-Sky Survey (2MASS) NIR catalog (Figure \ref{fig_cluster}a). $Chandra$ sources were considered to have stellar counterparts when the positional coincidences were better than $1\arcsec$ within $\sim 3.5\arcmin$ of ACIS-I field center, and $2\arcsec$ in the outer regions of each ACIS-I field where $Chandra$'s PSF deteriorates. Out of 225 X-ray sources, 190 have 2MASS matches. We have also acquired $JHK$ images of the central part of the cluster from the United Kingdom Infra-Red Telescope (UKIRT), which have better spatial resolution and sensitivity than the 2MASS images (Figure \ref{fig_cluster}b). Using UKIRT images 12 additional matches are identified, 10 of which are blended in 2MASS. 

Two of the three OB stars identified in \citet{Smith85} were matched to X-ray sources\footnote{These OB stars are sources \#99 and \#141 in Table \ref{tbl_phot}, which correspond to OS~2a and OS~1a in \citet{Smith85}. OS~1a is a blended source, where the components are distinguishable in the UKIRT image but not in the 2MASS image. The separation angle is 1.4$^{\prime\prime}$ and the position angle is 14$^{\circ}$. The position of the X-ray source corresponds to the position of the southern star.}. The third OB star was located on a chip gap and was not detected as an X-ray source. 

\citet{Rodriguez_in_prep} detect 20 Very Large Array (VLA) radio sources in the W40 complex near the center of the cluster, and 15 of them have X-ray counterparts. All sources with VLA counterparts also have infrared counterparts. The reported VLA source coordinates are rounded \citep{Rodney08}, so {\it Chandra}-VLA positional offsets may be greater than 1$^{\prime\prime}$ (\#141). Individual radio and NIR counterparts for {\it Chandra} sources are listed in Table \ref{tbl_irradio} along with their $JHK_s$ photometry. 

All X-ray sources with infrared counterparts are likely to be stars since YSOs have high apparent NIR fluxes compared to extragalactic contaminants. X-ray sources with NIR counterparts are classified as either W40 cluster stars or foreground stars (column 9 in Table \ref{tbl_irradio}). 

Most of the 23 X-ray detections without NIR counterparts are likely to be active galactic nucleii (AGN) \citep{Getman06}. However, some of these are classified as W40 members. Sources \#56, \#140, \#142, and \#155 have strong variability, which is a characteristic of T-Tauri X-ray emission, and source \#56 is located near the molecular core and is likely to be a highly embedded star. With their median energies $>$~3.5~keV, sources \#56 and \#155 are the most likely protostellar candidates, judging from the evolutionary class versus X-ray median energy relation shown in Figure~8 of \citet{Getman07}. 


\subsection{IR Properties of W40 Stars \label{ir_properties_section}}

\begin{figure}
\centering
\includegraphics[angle=0.,width=7.0in]{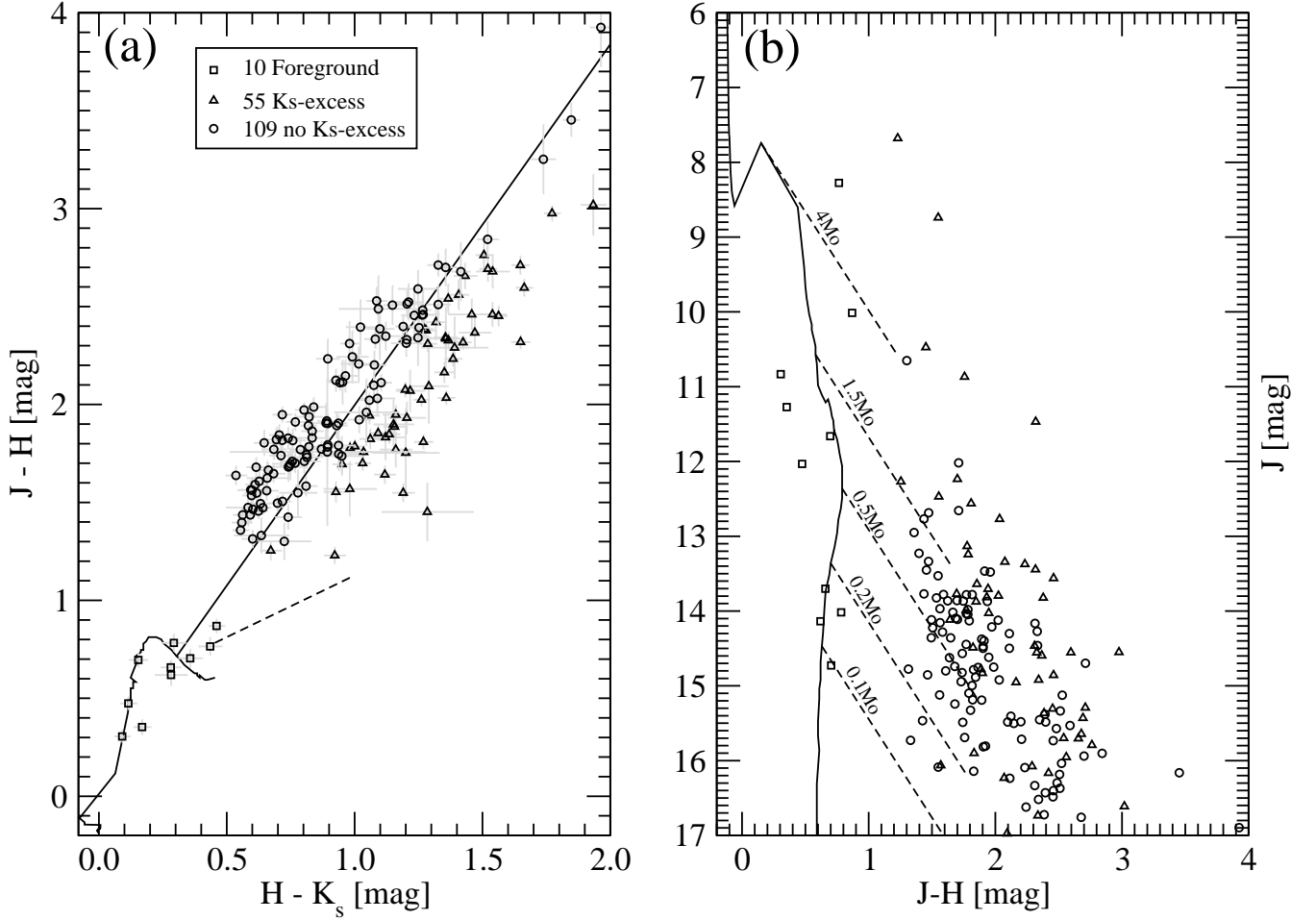}
\caption{NIR color-color (a) and color-magnitude (b) diagrams for 164 X-ray sources classified as cluster members with $K_s$ (triangle) and without $K_s$-band excess (circle), and foreground stars (square). In panel (a): the solid curved line shows sites of intrinsic $JHK_s$ colors of 1~Myr (see \S \ref{ir_properties_section}) stars; the straight solid line is a reddening vector (using the reddening law of \citet{Rieke85}) originating at $M=0.2$~M$_{\odot}$; the dashed line is the CTT locus \citep{Meyer97}. In panel (b) the solid line is 1~Myr isochrone, and the dashed lines are reddening vectors of $A_V = 10$~mag shown for various star masses. \label{fig_ir_properties}}
\end{figure}

Figure \ref{fig_ir_properties} presents NIR properties for most of the {\it Chandra} sources with NIR counterparts based on the photometric data in Table \ref{tbl_irradio}. The nature of sources with known NIR photometry can be estimated from these diagrams. Our analysis uses models of intrinsic $JHK_s$ photometric properties for stars with ages of 1~Myr assuming a distance of 600~pc (see \S\ref{section_XLF}). PMS evolutionary models of \citet{Baraffe98} and \citet{Siess97} are considered for the mass ranges of 0.02~M$_{\odot}\le M \le1.4~$M$_{\odot}$ and 1.4~M$_{\odot}\le M \le7~$M$_{\odot}$, respectively, and stellar photometric properties of stars with masses $>7$~M$_{\odot}$ are taken from \citet{Cox00}. Sources used in the analysis have errors on their NIR colors $<0.2$~mag.

The cluster stars are shifted to the upper right by $A_V\ge$5~mag on the color-color diagram (Figure \ref{fig_ir_properties}a), so the foreground stars are easily identified due to the lack of significant absorption. Ten foreground stars are detected.

The mass sensitivity limit of the X-ray detected PMS sample is $\sim0.2$~M$_{\odot}$; $\sim90$\% of stars have masses $<2$~M$_{\odot}$; and 7 stars have masses $\ga10$~M${\odot}$ (Figure \ref{fig_ir_properties}b). The 55 X-ray stars located $>$1.5~$\sigma$ to the right of the reddening vector originating at 0.2~M$_{\odot}$ (Figure \ref{fig_ir_properties}a) are classified here as $K_s$-band excess stars\footnote{Colors of massive stars identified as $K_s$-band excess stars are located to the right of the reddening vector originating at the location of intrinsic colors of massive stars (this reddening vector is not shown in Figure \ref{fig_ir_properties}).}, leaving 109 X-ray stars as non $K_s$-band excess stars. No stars with $K_s$-band excess are classified as protostellar objects; the X-ray observation is not sensitive to a protostellar population that might exist in W40 \citep{Bontemps10}. The disk fraction, taking into account the differences in X-ray sensitivity to stars with and without $K_s$-band excess, is estimated in \S \ref{section_XLF}.

Photometric mass and $V$-band absorption derived from the color-magnitude diagram (Figure \ref{fig_ir_properties}) are tabulated in Table \ref{tbl_irradio}. Uncertainties in the photometric mass estimates are discussed in Appendix \ref{mass_uncertainty_section}.  Typical uncertainties are around 0.2~M$_{\odot}$ for lower mass stars and 0.8~M$_{\odot}$ for higher mass stars. There is a degeneracy in $JHK_s$ photometry that leads to uncertain masses between 2~M$_{\odot}$ and 10~M$_{\odot}$ for the 1~Myr isochrone. We report the lower mass solution (2~M$_{\odot}-4$~M$_{\odot}$) as masses in this range are more likely due to the shape of the IMF.

\subsection{Membership \label{membership_section}}

Extragalactic sources, mostly AGNs at moderate redshifts, dominate X-ray source counts at high Galactic latitudes and some will be detected in fields near the Galactic plane despite the heavy obscuration \citep{Broos07}. Even the low X-ray detection rate of main-sequence stars will lead to contaminants in the X-ray image due to the huge number of foreground field stars. The field stars detected by X-ray observations will tend to be younger members of the disk population, since stellar X-ray emission declines rapidly after $\sim1$~Gyr \citep{Feigelson04,Preibisch05}. However, background field stars are a less important source of contaminating X-rays for W40, due to the high obscuration from the cloud. 

Out of 225 X-ray sources detected by our observation 19 are classified as extragalactic sources (\S \ref{identifications_section}) and 10 are classified as foreground stars (\S \ref{ir_properties_section}). The remaining 194 stars are likely W40 cluster members, although 30 do not have accurate IR colors for classification on the color-color diagram. Cluster membership is indicated in Table \ref{tbl_irradio}.

The foreground and extragalactic contamination of the W40 sources may be compared with the analysis of contamination in a {\it Chandra} ACIS-I observation of the young-star cluster Cep~B. The galactic latitude of Cep~B and W40 are somewhat similar ($\sim3.5\,^{\circ}$ for W40 and $\sim2.7\,^{\circ}$ for Cep~B), and the distances  (600~pc for W40 and 700~pc for Cep~B) and the {\it Chandra} exposures (38~ksec for W40 and 30~ksec for Cep~B) are similar, but the galactic longitudes (28.8$^{\circ}$ for W40 and 110$^{\circ}$ for Cep~B) are different. Confirmed by simulations, the numbers of extragalactic and foreground objects in Cep~B of 24 and 13, respectively \citep{Getman06}, are similar to those of W40.


YSC members are typically identified by IR excess, and $K_s$-band excess is used to detect PMS stars within the ACIS-I field of view missed by {\it Chandra}. The 38 additional 2MASS cluster members with $K_s$-band excess and the massive star OS~3a (no $K_s$-band excess) are shown in Table \ref{tbl_additional}. The $K_s$-band excess is identified in a similar way as for X-ray sources, but a reddening vector originating from 0.09~$M_{\odot}$ instead of 0.2~$M_{\odot}$ is used since the 2MASS point-source catalog is deeper than the {\it Chandra} observation and use of a reddening vector from 0.2~$M_{\odot}$ would likely find foreground diskless stars. None of these 39 sources have a VLA counterpart.

\section{X-ray Spectral Analysis \label{spectral_section}}

Low spectral resolution CCD spectra of young stars are typically modeled with one or two-temperature optically thin thermal plasma models subject to photoelectric absorption to measure flux, temperature, and hydrogen column density. The spectral analysis of X-ray stars is performed with the XSPEC spectral fitting package version 12.5 \citep{Arnaud96}. The unbinned source and background spectra are fit with one-temperature plasma emission models \citep[$apec$,][]{Smith01} using the maximum likelihood method \citep{Cash79}. We assume 0.3 times solar elemental abundances previously suggested as typical for young stellar objects in other star forming regions \citep{Imanishi01,Feigelson02}. Solar abundances are taken from \citet{Anders89}. X-ray absorption is modeled using the atomic cross sections of \citet[$wabs$,][]{Morrison83}.

X-ray luminosities and hydrogen column densities are also generated with XPHOT \citep[XPHOT program in][]{Getman10}. XPHOT is a non-parametric method for estimation of apparent and intrinsic broad-band fluxes and absorbing X-ray column densities of X-ray sources. Measured X-ray luminosity and absorption values from both XSPEC and XPHOT methods are in good agreement. The advantage of the XPHOT method is that it is more accurate than forward-fitting methods for very faint sources and provides both statistical and systematic (due to uncertainty in X-ray model) errors on derived intrinsic fluxes and absorptions. These errors on individual source X-ray luminosities are further used in our Monte-Carlo (MC) simulations to obtain errors on the X-ray luminosity functions of W40 stars (\S \ref{section_XLF}). 

The XSPEC and XPHOT spectral results are presented in Table \ref{tbl_spectroscopy}. The X-ray luminosities included for objects classified as non cluster members may become useful in cases of source misclassifications. In addition, the XPHOT luminosities for OB stars may not be accurate because the XPHOT method assumes the T-Tauri X-ray emission mechanisms \citep{Getman10}. 

\section{Confirmatory Results: X-ray Luminosity, Stellar Mass, and Absorption \label{confirmation_section}}

We would like to improve our confidence in the multiwavelength study of the W40 cluster by finding agreement between independently inferred X-ray and IR properties. Here the relationship between mass $M$ and X-ray luminosity $L_X$ and the relationship between visual extinction $A_V$ and X-ray column density $N_H$ are analyzed. 

The X-ray luminosity versus mass relationship was originally detected by \citet{Feigelson93} from {\it ROSAT} data. An empirical relationship of $L_X\propto M^{1.7}$, which extends over 3 orders of magnitude in $L_X$, can be seen in the COUP observation of the Orion Nebula Cluster \citep{Getman05a,Preibisch05b} and in the XMM-Newton Extended Survey of the Taurus Molecular Cloud (XEST) \citep{Gudel07a,Telleschi07}, respectively. This relationship has a poorly understood astrophysical cause, but may be an effect of the saturation of the magnetic dynamo in the fully convective interior or on the surface of PMS stars \citep{Preibisch05b}.

\begin{figure}
\centering
\includegraphics[angle=0.,width=7.0in]{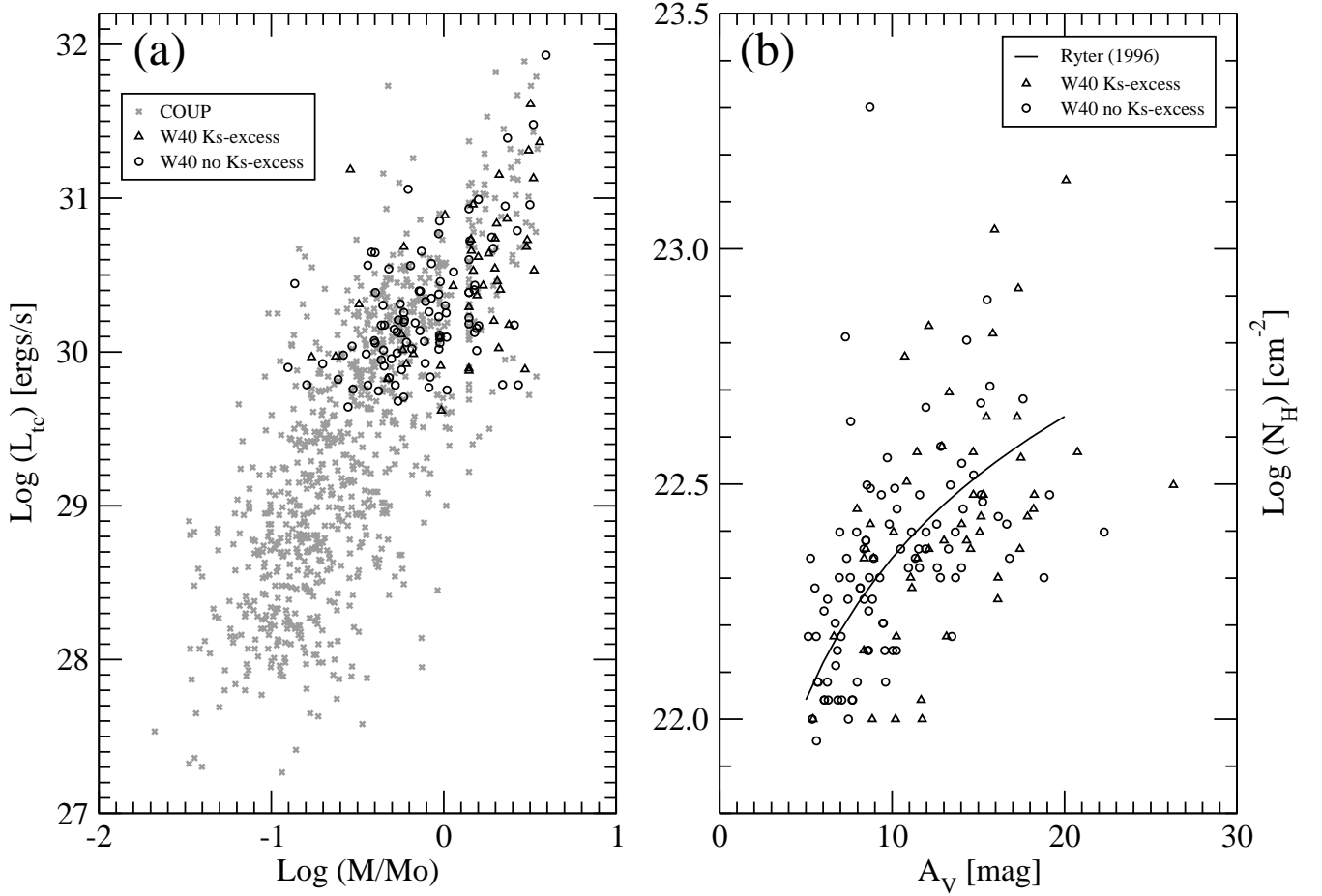}
\caption{(a) X-ray luminosity as a function of mass for lightly obscured PMS population in COUP (grey x), and W40 stars with (triangle) and without (circle) $K_s$-band excess. (b) X-ray absorbing column density from the X-ray spectral analysis with XPHOT and visual absorption from the IR photometry. The curve gives the gas-to-dust relationship $N_H = 2.2 \times 10^{21} A_V$ \citep{Ryter96}.  \label{fig_control}}
\end{figure}

W40 cluster stars are plotted on top of the sample of lightly obscured PMS stars from COUP on the $L_X$-$M$ diagram in Figure \ref{fig_control}a. For the W40 T-Tauri stars the X-ray luminosities are the total-band (0.5-8~keV) luminosities generated by XPHOT (\S \ref{spectral_section}), and the masses are derived from the NIR color-magnitude diagram (\S \ref{ir_properties_section}). The W40 stars occupy the same parameter space on this graph as the COUP stars, down to the sensitivity limit of the {\it Chandra} observation of W40. 

IR absorption is sensitive to dust while bound-free absorption of X-ray photons measures the integrated effects of interstellar material in any phase (partially ionized, neutral, molecular, solid). Contributions to the X-ray absorption come from elements including He and inner shell electrons in C, N, O, Ne, Si, S, Mg, Ar, and Fe atoms \citep{Wilms00}, and the absorption is expressed as a hydrogen column density. The relation found between $N_H$ and $A_V$ reveals the gas-to-dust ratio for a cluster. The relation between $N_H$ and $A_V$ in W40 (shown in Figure \ref{fig_control}) is in agreement with the gas-to-dust ratio of the galactic interstellar medium \citep{Ryter96}. Both samples of stars with and without $K_s$-band excess follow the trend: $\log N_H / \log A_V = 21.30\pm0.03$ for stars with $K_s$-band excess, and $\log N_H / \log A_V = 21.34\pm0.02$ for stars without $K_s$-band excess. Both the $L_X-M$ and $N_H-A_V$ agreements provide confidence in our X-ray and NIR analyses. 

\section{XLF \label{section_XLF}}

\begin{figure}
\centering
\includegraphics[angle=0.,width=4.5in]{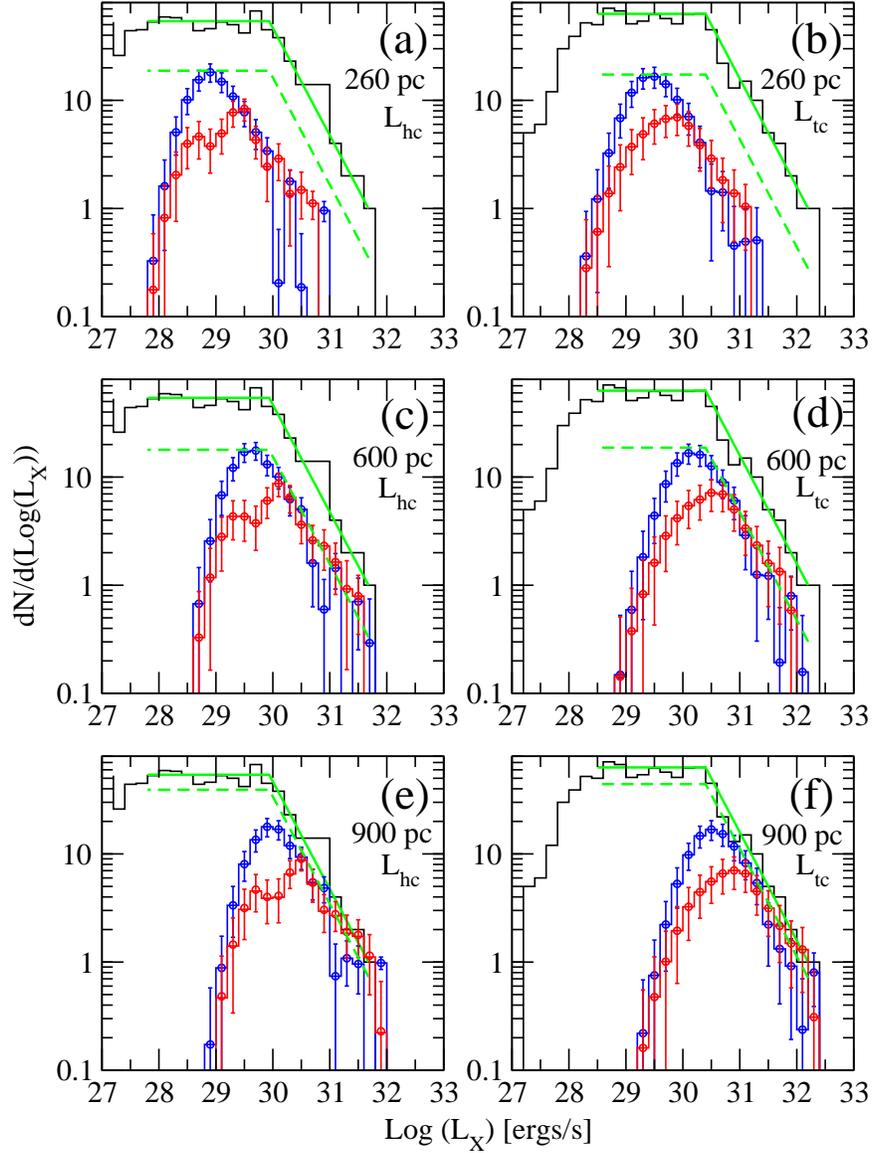}
\caption{Comparison between XLFs of the ONC and W40 populations. The upper black histograms shows the COUP unobscured cool sample of $>800$ stars \citep{Feigelson05}; the red and blue histograms show samples of W40 stars with and without $K_s$-excesses, respectively. Error bars indicate 68\% confidence intervals (1~$\sigma$) from Monte Carlo simulated distributions when X-ray luminosities are randomly drawn from Gaussian distributions with the source's measured $\log L_X$ values and estimated errors $\Delta \log L_X$. The green lines are added to aid the eye; they are based on the shape of the ONC XLF and scaled downward to match the W40 population without $K_s$-excess. Hard-band and total-band XLFs are shown on the left and right, respectively. Panels (a)-(b), (c)-(d), and (e)-(f) are for the assumed distances to W40 of 260, 600, and 900~pc, respectively. \label{fig_xlf}}
\end{figure}

The statistical link between X-ray luminosity and stellar mass allows an association between the XLF and the IMF. A universal XLF has been hypothesized based on COUP and observations of IC~348 and NGC~1333 \citep{Feigelson05,Feigelson05a}. The assumption of a universal XLF has been used to estimate total populations of M17 \citep{Broos07}, Cep~B \citep{Getman06}, NGC 6357 \citep{Wang07}, and NGC 2244 \citep{Wang08}. NGC 2244 and Cep~B are complete beyond the COUP XLF's turnover at $\log L_{hc}=30.0$~erg~s$^{-1}$ (Figure \ref{fig_xlf}), and while the NGC 2244 XLF closely matches the COUP XLF, there is a slight inconsistency between the Cep~B XLF and the COUP XLF due to the excess of stars with $\log L_{hc}=29.3$~erg~s$^{-1}$ in Cep~B. X-ray luminosity is known to decline during main-sequence evolution, likely causing the XLF of a cluster to change. However, The XLF does not strongly evolve during the PMS era for ages $\la 5$~Myr \citep{Preibisch05, Prisinzano08}. The assumption of a universal XLF can also be used for YSC distance measurements \citep{Feigelson05a}. For example the XLF analysis for Serpens Main cluster by \citet{Winston10} favors the distance of 360~pc over 260~pc, which is closer to the recent VLBA parallax measurement of 415~pc \citep{Dzib10}.

The W40 XLFs are simulated by varying reported XPHOT luminosities of individual sources based on their errors, similar to the analysis of \citet{Getman06}, and are compared with the COUP XLFs of $>800$ unobscured stars \citep{Feigelson05}. Shifting the W40 XLF horizontally corresponds to changing the distance to W40, and the vertical shift of the COUP XLF model corresponds to scaling COUP to the W40 population. 



X-ray observations are less sensitive to stars with circumstellar disks \citep{Getman09}, so we divide the stars into groups based on the presence of $K_s$-band excess; Figure \ref{fig_xlf} shows W40 XLFs of 50 stars with $K_s$-band excess and 96 stars without $K_s$-band excess in the hard and total bands. Both populations have similar XLF distributions above $L_{hc}=10^{30.0}$~erg~s$^{-1}$ (assuming $D=600$~pc), but differ at lower X-ray luminosities. We also perform XLF analysis on the entire sample of 189 cluster members that have available X-ray luminosity information without dividing the sample based on $K_s$-band excess (graph not shown). 

\subsection{Distance \label{section_distance}}

In Figure \ref{fig_xlf} the COUP XLF is compared to the W40 XLF with three different trial distances: 600~pc, the most commonly reported distance; 260~pc, the distance to the nearby on the sky Serpens South cluster; and 900~pc, the largest reported distance (\S \ref{introduction_section}). Distances of 600 and 900~pc provide good fits, with a less complete 900~pc observed XLF. A distance of 500~pc provides a qualitatively worse fit, with a slight excess of stars above the COUP XLF at a $\log L_{hc}\sim 30$~erg~s$^{-1}$ (graph not shown). With the assumption of a universal XLF, the W40 XLF analysis favors the distance of $600-900$~pc; throughout the paper we adopt the most often reported distance of 600~pc.

Further on we carry out the analysis assuming $d=600$~pc. However, as mentioned in \S \ref{past_section}, some studies favor the distance of 260~pc. Therefore, below we comment on possible modifications of the results if a distance of 260~pc is considered instead.

If the distance to W40 is 260~pc, then the excess of stars with $28.5<L_{hc}<30$~erg~s$^{-1}$ is inconsistent with a universal XLF. This may indicate the absence of the expected higher mass stars in the cluster (see \S \ref{imf_section} for a discussion of the IMF for 260~pc), or may be due to $L_X$ evolution if the cluster is much older than 1~Myr. The excess would be comparable to the excess of stars at $L_{hc}=29.3$~erg~s$^{-1}$ in Cep~B \citep{Getman06}.

\subsection{Completeness Limits, Total Population, Disk Fraction, and Age \label{tot_pop_section}}

The X-ray completeness limits, i.e. where the W40 XLF diverges from the scaled COUP model, are $\log L_{hc}=30.2$ and $\log L_{tc}=30.6$ corresponding to 50\% completeness at 1.4~M$_{\odot}$.


The vertical shift between the COUP and the W40 XLFs for the total as well as the $K_s$-band excess stratified star samples give the same estimate of the total W40 stellar population down to 0.1~M$_{\odot}$ of 70\% the COUP sample, i.e. 600 PMS stars (Table \ref{table_xlf_completeness}). 

The same inferred vertical scaling of 35\% to fit the COUP XLF model to both the W40 XLFs of stars with and without $K_s$-band excess gives a W40 $K_s$-band excess fraction of $\sim$50\%. The $K_s$-band excess fraction found for the W40 cluster is likely to be a lower limit on the disk fraction, since $JHK$ is not the most sensitive tracer of disks. For example, $L$-band photometry in addition to $JHK$ gives a more robust estimate on the fraction of stars with circumstellar disks. Clusters such as IC 348 \citep{Haisch01b} (age $\sim3$~Myr), NGC 2024 \citep{Haisch01c} (age $< 1$~Myr), and the Orion Nebula Cluster \citep[ONC;][]{Lada00} (age $1-2$~Myr) have the following reported $JHK$ ($JHKL$) disk fractions: $20-25$\% (65\%) in IC348, 60\% (85\%) in NGC 2024, and $55-90$\% (85\%) in the ONC. The $JHK_s$ disk fraction of 50\% thus suggests that the W40 cluster is young with an age of likely $\le1$~Myr. W40's likely astrophysical disk fraction of $\sim$80\% gives an age of 0.5~Myr using the regression line in \citet{Mamajek09} for age versus disk fraction of 22 star clusters.

However, there is a somewhat controversial situation, since 6 out of the 8 high-mass star candidates ($M\ga10$~M$_{\odot}$) have $K_s$-band excess, suggesting a $K_s$-band excess fraction of $75^{+9}_{-19}$\%\footnote{Errors on disk fractions have been estimated using binomial distribution statistics as described by \citet{Burgasser03}.}. Theoretical disk depletion timescales for early-type stars, due mostly to photoevaporation with some contribution from material removal by winds, are roughly 0.2~Myr and 0.7~Myr for O7V and B1V stars, respectively \citep{Hollenbach94,Hollenbach00}. The observed disk fractions and the theoretical depletion timescales might suggest that most of W40 X-ray star high mass candidates formed somewhat later than low mass stars, for example the latter phenomenon is reported for stars in W3 \citet{Feigelson08}, M17 SWex \citet{Povich10}, and IRAS 19343+2026 \citet{Ojha10}. 

If W40 were at 900~pc the XLF would indicate completeness limits at $\log L_{hc}=30.4$~erg~s$^{-1}$ and $\log L_{tc}=30.8$~erg~s$^{-1}$ and a total population of at least 1000 stars down to 0.1~M$_{\odot}$, but the disk fraction would not change. However, if the distance is 260~pc then these values cannot be derived from the XLF since the shape of the XLF would differ from the shape of the COUP XLF.

\section{IMF \label{imf_section}}

\begin{figure}
\centering
\includegraphics[angle=0.,width=5.5in]{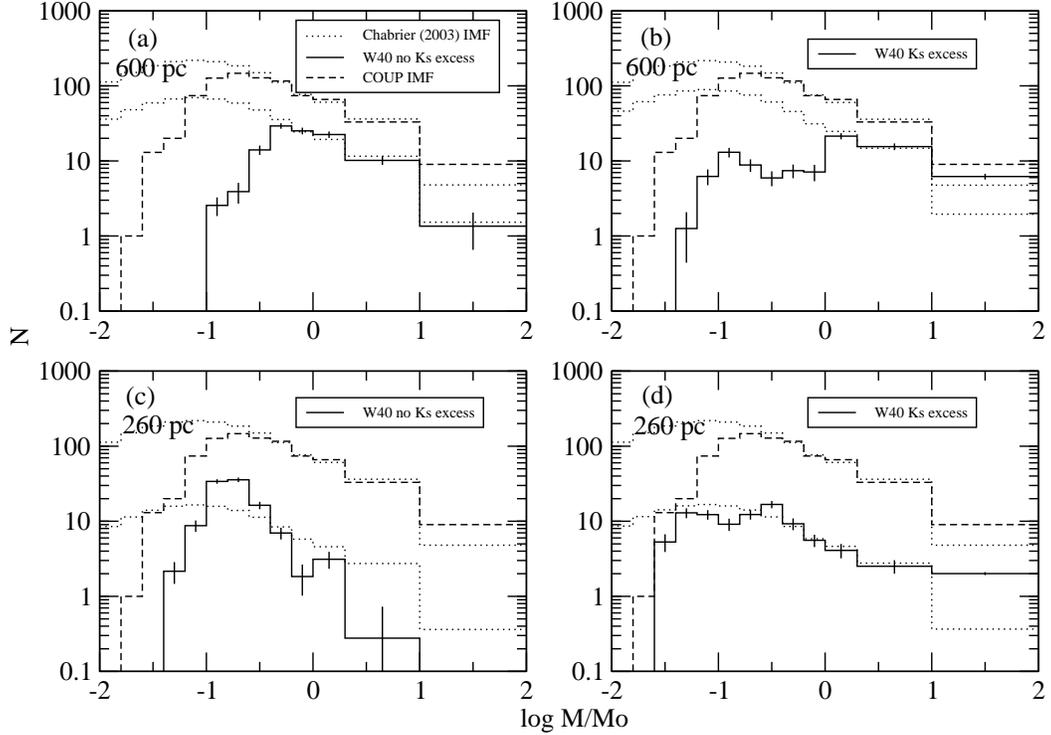}
\caption{Comparison between IMFs of the COUP (dashed lines) and the W40 populations (solid lines). The top two panels show IMFs for 600~pc for stars without $K_s$-band excess (a) and with $K_s$-band excess (b). The bottom two panels show IMFs for 260~pc for stars without $K_s$-band excess (c) and with $K_s$-band excess (d). Bin sizes are adjusted to cover the entire region of $JHK_s$ photometric mass degeneracy with a single bin (\S \ref{ir_properties_section}), and 1~$\sigma$ simulated errors are shown. The \citet{Chabrier03} IMF (dotted lines), integrated to match the binning, is scaled to the COUP and W40 IMFs.\label{fig_imf}}
\end{figure}

To verify the results of the XLF analysis, the IMF of the cluster is inspected (assuming a distance to W40 of 600~pc). For consistent treatment, in the IMF analysis the stellar masses used for both W40 and COUP are the masses derived by the method described in \S \ref{ir_properties_section}. The W40 X-ray and IR selected samples are combined and compared to the sample of $>800$ unobscured COUP stars. The \citet{Chabrier03} IMF is scaled to the COUP IMF, the IMF of sources without $K_s$-band excess, and the IMF of sources with $K_s$-band excess (Figure \ref{fig_imf}ab). Error bars on the W40 IMF are generated through simulation\footnote{The simulation takes into consideration effects of statistical errors, systematic errors, and mass degeneracy for some stars. The selection of stars in each mass bin is generated 1000 times with individual mass values calculated from $JHK_s$ photometry where magnitudes are drawn from a Gaussian distribution with a standard deviation equal to error reported by 2MASS. For stars with $M<5$~M$_{\odot}$, masses are drawn from a Gaussian distribution with a mean equal to the mass derived above and a standard deviation of 30\% to account for systematic error found in Appendix \ref{mass_uncertainty_section}. For stars with multiple mass solutions, a solution is randomly chosen with a weight based on the \citep{Chabrier03} IMF. However, the effect on the IMF is minimized since a single mass bin spans the entire region of degeneracy. }.

W40 cluster members with no $K_s$-band are complete to 0.5~M$_{\odot}$ and stars with $K_s$-band excess are complete to 1.5~M$_{\odot}$. This agrees with the completeness result from the XLF analysis. From the scaling of the \citet{Chabrier03} IMF, the population of stars without $K_s$-band excess (with $K_s$-band excess) is 32\% (40\%) the size of the COUP sample, implying a $JHK_s$ disk fraction of 55\% and a total population of $\sim$600 PMS stars down to 0.1~M$_{\odot}$ (Table \ref{table_xlf_completeness}). These results are consistent with those of the XLF analysis.

If W40 is at a distance of 260~pc, the IMF for both stars with and without $K_s$-band excess would be complete down to 0.1~M$_{\odot}$, so the total cluster population would be $\sim$200 stars (Figure \ref{fig_imf}cd). The disk fraction would be $\sim$50\%. However, at this distance the IMF for stars without $K_s$-band excess has an excess of stars between 0.1 and 0.3~M$_{\odot}$, so the IMF does not resemble either the COUP IMF or the \citet{Chabrier03} IMF. This result also indicates that 260~pc is too near a distance for W40.  




\section{Massive Star Candidates \label{section_OB}}

W40 is known to contain massive stars, including three OB stars that act as ionizing sources for the cluster \citep{Smith85}. Eight cluster members have photometric masses $\ga10$~M$_{\odot}$ (Table \ref{OB_table}), and the masses of identified ionizing sources OS~1a, OS~2a, and OS~3a are consistent with OB stars. OS~1a is a visual double and only binary IR photometric properties are known. OS~2a has a mass of 30~M$_{\odot}$ and OS~3a (which is not observed in the X-ray due to location on the chip) has a mass of 15~M$_{\odot}$. Out of the 8 high mass star candidates, 6 have $K_s$-band excess, including OS~1a and OS~2a, but not OS~3a. \citet{Smith85} detect circumstellar material around OS~1a, OS~2a, and OS~3a, and \citep{Vallee94} find evidence for substantial circumstellar material around OS~1a and OS~2a. 

However, if a distance of 260~pc is considered only two stars, OS~1a and OS~2a, have $JHK_s$ masses $\ga10$~M$_{\odot}$.

X-ray properties inferred from one-temperature spectral fits\footnote{Results are given for one-temperature thermal fits ($wabs*apec$). To account for mild pile-up in source \#122 the model $pileup$ from \citet{Davis01} is applied. Reduced $\chi^2$ is used to fit models to all the sources, except for sources \#46 and \#144 which have a small number of counts and are fit using the C-statistic \citep{Cash79} on the unbinned spectrum.} are also given in Table \ref{OB_table}. We caution that due to the high absorption of the W40 cluster, it is difficult to infer the existence of very soft temperature components. In addition lack of low energy photons may bias temperatures inferred by XSPEC to higher values.

Soft ($\la 1$~keV) and nearly constant X-ray emission from O and early-type B stars is often explained by a wind-shock model, where shocks form in a freely expanding wind subject to the Line-Driven Instability \citep{Lucy80,Owocki99}. Hard and often variable X-ray emission from massive stars is proposed to be produced by either the wind-wind collision mechanism in a massive binary system \citep{Zhekov00} or the Magnetically Channeled Wind Shock Mechanism \citep{Babel97,Gagne05}. None of the high mass star candidates is X-ray variable. Considering large uncertainties of $\sim 50$\% on inferred plasma temperatures, spectra of two sources, \#122 and \#145, are significantly inconsistent with the purely soft ($\la 1$~keV) plasmas of the classical Line-Driven Instability model. Highly ionized spectral lines of Mg, Si, and S with formation temperatures above 2~keV are also seen in the spectrum of source \#122, suggesting the presence of the hot plasma component. An alternative to the wind-wind collision or Magnetically Channeled Wind Shock Mechanism explanations is that the hard X-ray emission is not coming from the OB stars themselves, but from unresolved T-Tauri companions.

\begin{figure}
\centering
\includegraphics[angle=0.,width=6.0in]{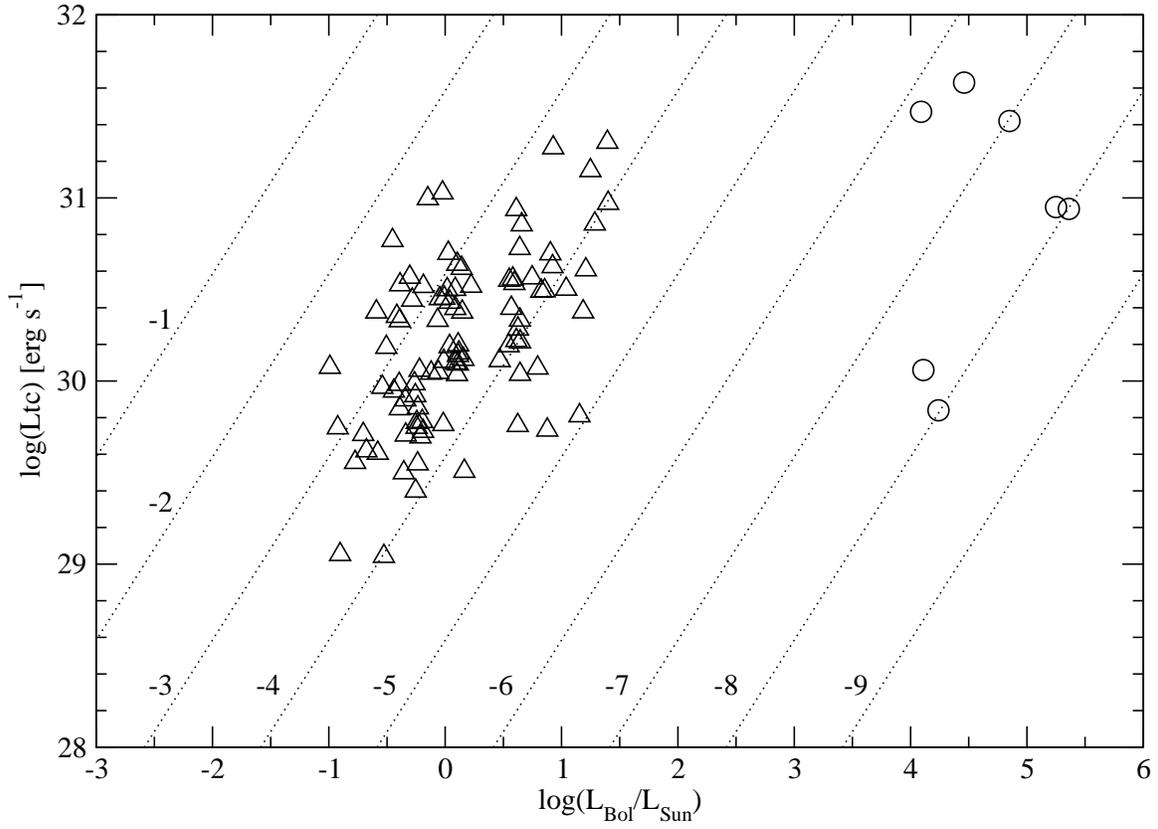}
\caption{Total-band (0.5-8.0~keV) X-ray luminosity as a function of bolometric luminosity. Massive star candidates are circles and lower mass stars are triangles. Dotted lines indicate $L_X/L_{\rm Bol}=10^{-1},10^{-2},10^{-3},10^{-4},10^{-5},10^{-6},10^{-7},10^{-8},$ and $10^{-9}$.\label{lx_vs_lbol_fig}}
\end{figure}

The level of X-ray emission of O and early-type B stars is often comparable to or higher than the most X-ray luminous T-Tauri stars \citep[e.g.][]{Stelzer05}. The OB star emission roughly follows the trend $L_{\rm X}=10^{-7}L_{\rm Bol}$ \citep{Chlebowski89,Berghoefer97}. Figure \ref{lx_vs_lbol_fig} shows the $L_{tc}-L_{\rm bol}$ relation for the W40 T-Tauri stars and OB candidates. Bolometric luminosities are derived from the NIR color-magnitude diagram (\S \ref{ir_properties_section}).The mean $\log(L_{tc}/L_{\rm Bol})$ for W40 OB candidates is -7.3 and the distribution has a standard deviation of 0.7. This is similar to the relation found for other clusters observed by {\it Chandra}, for example M17 \citep{Broos07} and NGC 2244 \citep{Wang08}. However, the X-ray emission level of some W40 high-mass candidates may be explained by an X-ray active T-Tauri companion unresolved in the NIR.

\section{Spatial Structure \label{spatial_section}}
 
\subsection{General Cluster Properties \label{section_general_spatial}}

For the analysis of W40's structure, we choose to define the center of the cluster as the median position of all {\it Chandra}-detected members, which has coordinates R.A.$=$18$^{\rm h}$ 31$^{\rm m}$ 25.\hspace{-0.6ex}$^{\rm s}$14, decl.$=$-2$^{\circ}$ 05$^{\prime}$ 35.\hspace{-0.6ex}$^{\prime\prime}$7 (J2000). An alternative strategy is to use the OB stars as the center of the cluster, but in W40 the median position of the high mass stars is offset from the median positions of other stellar populations as shown in Table \ref{structure_table}, so we judge that the median position of all the stars is a better estimate. For discussion of projected radial distances in the cluster $d=(D/600~$pc$)$ accounts for uncertainty in the distance $D$ to W40.

\begin{figure}
\centering
\includegraphics[angle=0.,width=6.0in]{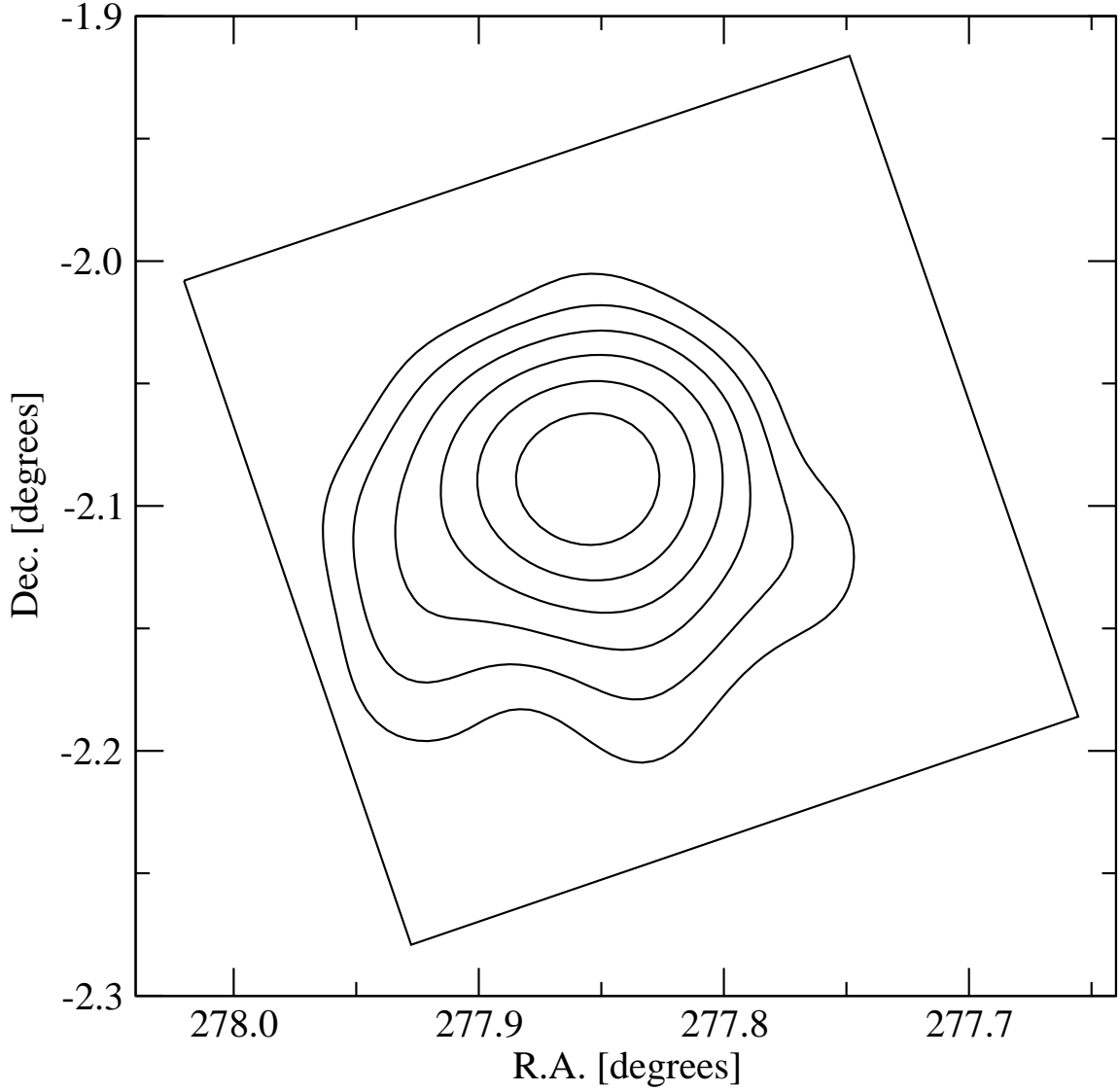}
\caption{The surface density of W40 cluster members detected in the X-ray, smoothed with a 1.5~arcmin Gaussian kernel. Contours are drawn starting at 100 stars degree$^{-2}$ and at every subsequent increase by a factor of 1.5 in surface density.\label{fig_spatial_contour}}
\end{figure}

Fifty percent of the {\it Chandra}-detected cluster members are within 0.50$d$~pc of the cluster center and 75\% are within 0.90$d$~pc. The edge of the ACIS-I array is 7.5$^{\prime}$ (1.3$d$~pc) from the center of the cluster, so our analysis of structure is restricted to the region within this radius. The most distant detected star is 1.8$d$~pc from the center. The core radius is $\sim$0.15$-$0.2$d$~pc (\S \ref{section_profile}). In the core there is a small clump of 8 stars, including OS~1a, within a radius of 10$^{\prime\prime}$. A smoothed stellar surface density plot of the cluster is shown in Figure \ref{fig_spatial_contour}.


The structure of YSCs may be affected by both the locations where stars form and by cluster dynamical evolution. The velocity dispersion of stars in W40 is likely to be similar to other YSCs; \citet{Furesz08} find a radial velocity distribution of $\sigma=3.1$~km s$^{-1}$ in the ONC, and simulations of small and large clusters find one-dimensional stellar velocity dispersions of 1.2~km s$^{-1}$ and 4~km s$^{-1}$ \citep{Bate03,Bate09}. Assuming the stellar velocity dispersion in W40 of $\sim$2-4 km s$^{-1}$ and a cluster size of 2$d$~pc, the cluster crossing time is $t_{{\rm cross}}\sim0.5-1d$ Myr, on the order of the age of the cluster. The two-body relaxation time for a cluster is 
\begin{equation}t_{{\rm relax}}=\frac{N}{8\ln N}\times t_{{\rm cross}},\end{equation}
where N is the number of stars \citep{Binney87}. In W40, N$\sim$600 so the two-body relaxation time is $\sim5-10$ Myr, longer than the age of the cluster. However, violent relaxation may proceed more quickly \citep{Lynden-Bell67}. 

If the distance of 260~pc is considered instead of 600~pc, the relaxation time would decrease to $\sim1-2$~Myr.

\subsection{Surface Density Profiles\label{section_profile}}

\subsubsection{Radial Profile \label{radial_profile}}

\begin{figure}
\centering
\includegraphics[angle=0.,width=6.0in]{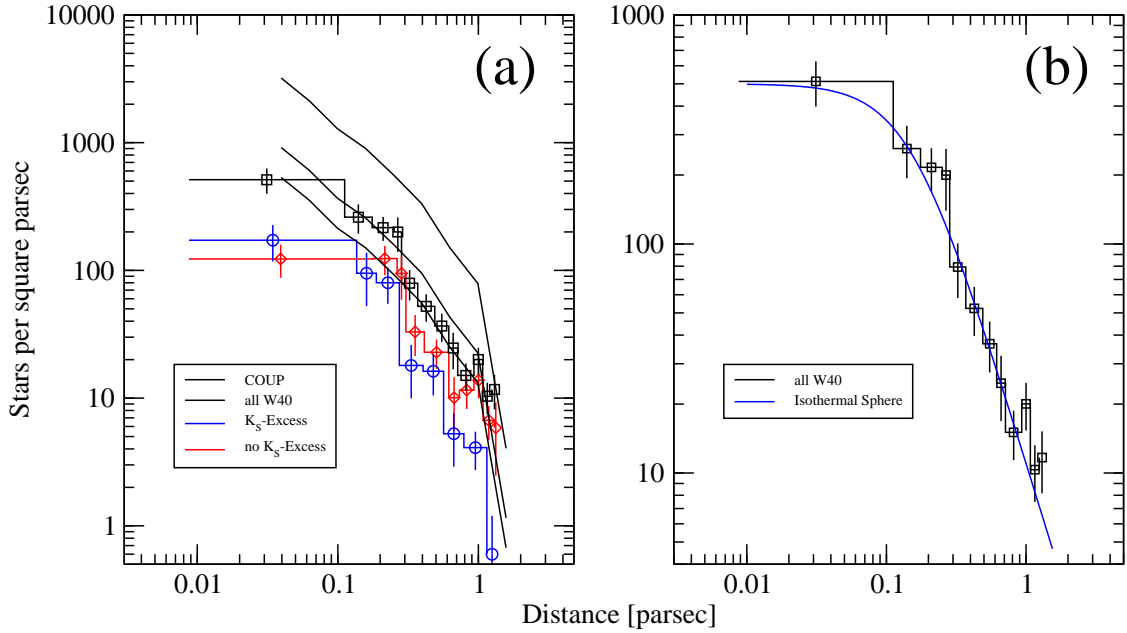}
\caption{(a) The radial profile of all W40 cluster members (black squares), stars with $K_s$-band excess (blue squares), and stars with no $K_s$-band excess (red squares). $\sqrt{N}$ error bars are used. On top, the radial profile of stars in COUP is plotted for comparison (black solid line), and it is scaled to match the various W40 radial profiles. (b) The radial profile of all W40 cluster members is shown in black, and the best King profile \citep{King62} found from fitting the cumulative distribution is shown in blue. \label{fig_radial_profile}}
\end{figure}

Figure \ref{fig_radial_profile} shows the radial surface density profile of W40. All cluster members detected by {\it Chandra} are included to improve counting statistics. The radial profile of stars from COUP \citep{Feigelson05} is provided for comparison, and a scaled version of the COUP radial profile closely resembles the radial profile of W40. Various functional profile models are evaluated using the cumulative distribution function (CDF) of the radial distances of cluster members and the K-S statistic. No single power law provides a good fit. However, the inner 0.2$d$~pc part of the profile is flatter and can be fit with a power law index of -0.2, and the outer part of the cluster is fit with a power law index of -1.6. The 90\% confidence interval on the indices is $\pm$0.3. YSC surface-density profiles are often fit by shallower power laws that typically have an exponent of -1. Examples include IC 348 ($\sigma\propto r^{-1}$) and NGC 2282 ($\sigma\propto r^{-1.1}$) \citep{Lada95,Horner97}. 

Assuming spherical symmetry, a surface density distribution with a power law $\sigma\propto r^{\alpha}$ and $\alpha<0$ will correspond to a three-dimensional density distribution with a power law $\rho\propto r^{\alpha-1}$, so the three-dimensional power law index for the inner part of the cluster will be approximately -1.2 and the power law index for the outer part of the cluster will be -2.6 (similar to the three-dimensional power-law index derived from the $Q$ parameter analysis described in \S \ref{Q_section}). 

We define the core of W40 as the region within 0.2$d$~pc, since this region has a noticeably less steep profile than the outside of the cluster. Within the core there are 48 {\it Chandra}-detected stars, the three ionizing sources OS1a, OS2a, OS3a, and 3 other high mass star candidates. This core radius is similar to the core radius of the ONC which was found to be $r_c\simeq0.15$~pc \citep{Hillenbrand98}.
 
The ONC stellar density profile is well fit by the King profile \citep{Hillenbrand98} for an isothermal sphere \citep{King62} that is applicable to gravitationally relaxed clusters and is described by the equation
\begin{equation}\sigma(r)\approx\sigma_c(r)=\frac{\sigma_0}{1+(r/r_c)^2}.\end{equation}
Nevertheless, \citet{Hillenbrand98}, \citet{Feigelson05}, and \citet{Furesz08} argue that the Orion cluster is not truely dynamically relaxed. The W40 cluster can also be fit by an isothermal sphere with a core radius of 0.15$d$~pc (K-S probability of 92\%), but W40 might also not be dynamically relaxed due to its young age (\S \ref{section_general_spatial}). 

\subsubsection{Profile Fitting and Detection of Subclustering with the $Q$ Statistic\label{Q_section}}

\begin{figure}
\centering
\includegraphics[angle=0.,width=6.0in]{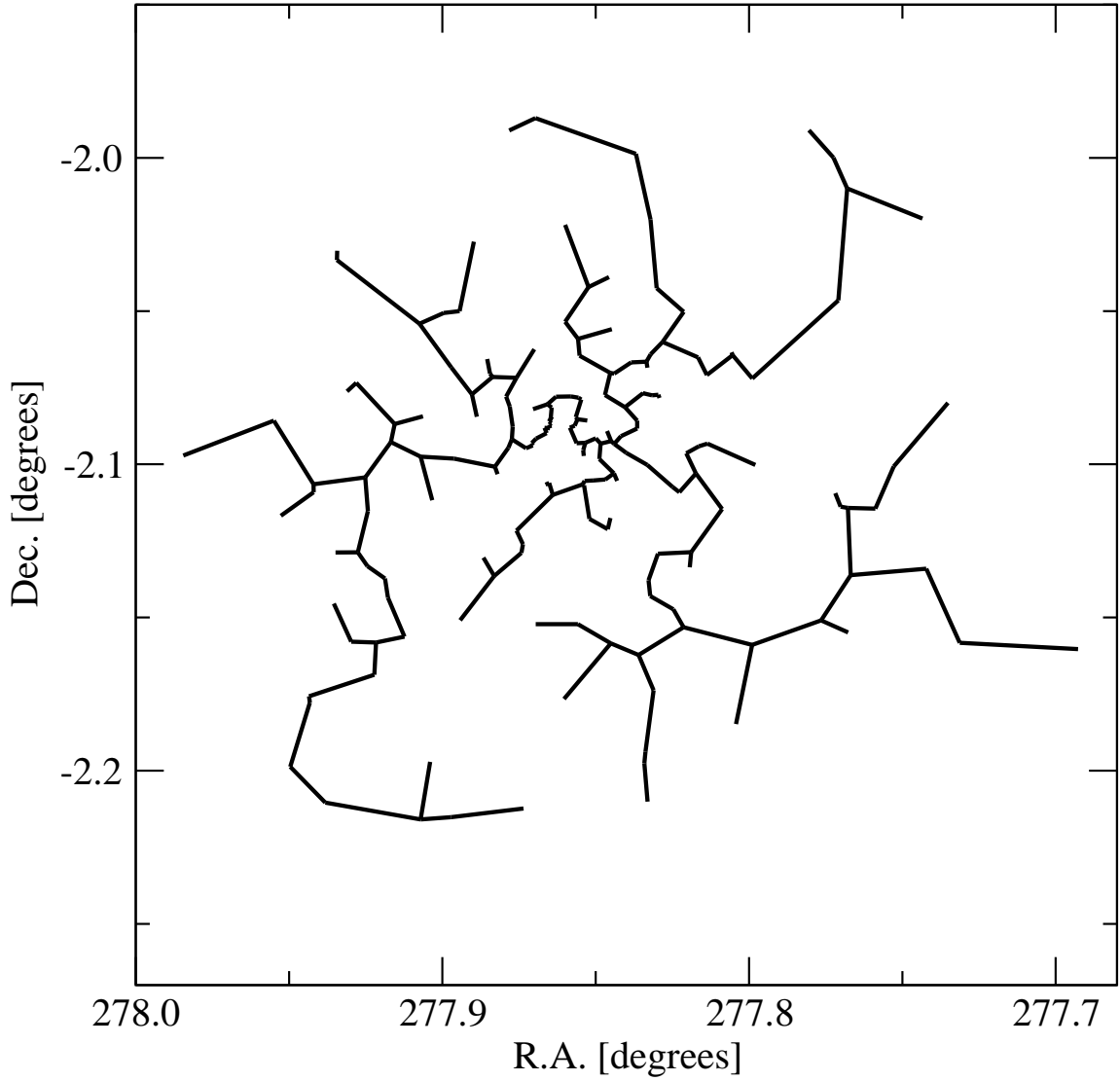}
\caption{The minimal spanning tree of W40. {\it Chandra}-detected cluster members are vertices. \label{fig_mst}}
\end{figure}

A statistic $Q$ indicates the existence of subclustering and measures the central concentration of clusters without subclustering \citep{Cartwright04}. This parameter combines a measure of the separation between stars and a measure of the edge lengths of the Euclidean minimal spanning tree (EMST). The EMST is the unique graph with minimum total edge length that connects all vertices. Figure \ref{fig_mst} shows the EMST for all W40 cluster members detected by {\it Chandra}, where the stars are the vertices of the graph. The parameter $s$ is the mean separation between stars, and $m$ is the mean edge length on the graph. \citet{Cartwright04} define
\begin{equation}\bar{s}=s/r_{{\rm max}}\end{equation} and
\begin{equation}\bar{m}=\frac{m(N_{{\rm max}}-1)}{N_{{\rm max}}\pi r_{{\rm max}}^2} ,\end{equation}
where $r_{{\rm max}}$ is the maximum radius, which is 7.5$^{\prime}$ for W40 (stars outside of this radius are excluded), and $N_{{\rm max}}$ is the number of sources within the radius. The parameter $Q$ is defined by
\begin{equation} Q=\frac{\bar{m}}{\bar{s}}. \end{equation}
Figure 5 in \citet{Cartwright04} gives fractal dimension or central concentration as a function of $Q$ generated from simulated clusters. A YSC with subclustering will have a value of $Q<0.8$. For W40 $Q=1.47$, indicating that W40 does not have subclustering (within 7.5$^{\prime}$) and that the cluster has a three-dimensional density profile $\rho\propto r^{\alpha}$, where $\alpha=-2.7$. In addition, a visual examination of the EMST in Fig \ref{fig_mst} reveals no unusually long segments, which suggests that no subclustering is present. The analysis also agrees closely with the results of the power law fits to the outer part of the cluster (\S \ref{radial_profile}). For comparison, the shallower three-dimensional radial profile exponents of -1.2, -2.2, -1.7 are found for $\rho$ Ophiuchus, IC348, and $\sigma$ Ori \citep{Cartwright04,Caballero08}, respectively.


\subsubsection{Cluster Elongation and Asymmetry \label{section_elongation}}

There is no evidence of that the W40 cluster of all {\it Chandra}-detected PMS stars is elongated; no correlation is found between right ascension and declination of cluster members (Person correlation coefficient $=$ -0.1), and the standard deviations of right ascensions and declinations are similar (173$^{\prime\prime}$ and 180$^{\prime\prime}$, respectively).

\begin{figure}
\centering
\includegraphics[angle=0.,width=6.0in]{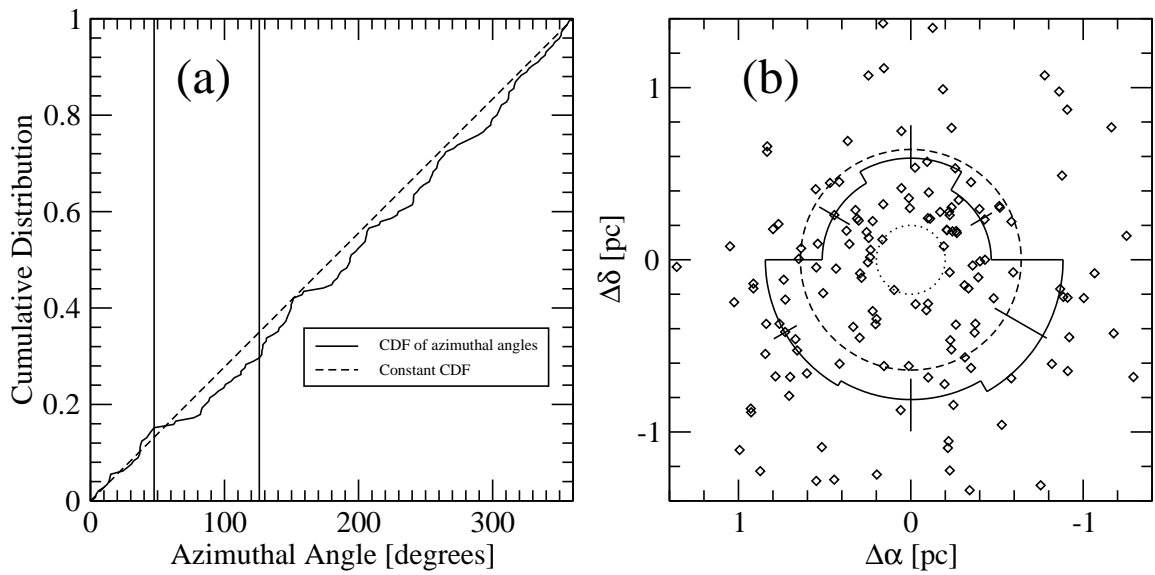}
\caption{(a) The cumulative distribution function of azimuthal angles. (b) The solid line marks the median distance of stars from the cluster center in each 60~degree segment. Error bars are plotted, and the best fitting constant radius is shown (dashed line). The boundary of the center of the cluster from which stars are excluded is shown with a dotted line. Stars are indicated by diamonds. \label{radial_symmetry_fig}}
\end{figure}

To further investigate whether the exterior of W40 ($r>0.2d$~pc) has radial symmetry, the distribution of azimuthal angles of stars is compared to an isotropic distribution, and the median distances of sources from the center is examined as a function of azimuthal angle (Figure \ref{radial_symmetry_fig}). For the former test the CDF of the angles was found to be consistent with a uniform distribution using Kuiper's statistic \citep{Kuiper62}, providing additional evidence that subclustering does not exist in W40. For the latter test, sources are binned by azimuthal angle with bin sizes of 60$^{\circ}$, and median distances are determined with error on the median calculated using 1000 bootstrap values. The median distances are consistent with a single value (reduced $\chi^2$ is 0.4 when fit with a constant). However, with bin sizes of 180$^{\circ}$ the reduced $\chi^2$ is 3.4 due to larger radial distances in the south half of the cluster, indicating that these stars are less centrally concentrated.  


Most YSCs, including both the ONC \citep{Hillenbrand98} and $\sigma$ Orionis \citep{Caballero08} have asymmetric morphologies. These asymmetric morphologies have been taken as another sign of incomplete dynamical relaxation \citep{Hillenbrand98} and may be an effect of the elongation of the star-forming molecular clouds \citep{Allen07}. If W40 is not dynamically relaxed, the near radial symmetry could be a result of spherically symmetric star formation or a projection effect if we are looking down the asymmetric axis of the cluster. \citet{Furesz08} caution about the interpretation of radial profiles that average over all angles, since asymmetry may cause steeper power laws, but this is not a great concern for W40 because of the near radial symmetry.

Finally, if considering only sources with $K_s$-band excess the azimuthal distribution is not uniform (0.5\% K-S probability). We further study this in \S \ref{section_ksdistribution}.


\subsection{Mass Segregation \label{section_mass_segregation}}

The dependence of cluster structure on stellar mass in YSCs provides clues about cluster formation and dynamical evolution. It is observed that massive stars are preferentially detected near the centers of open clusters and YSCs. Mass segregation is an expected result of dynamical relaxation, but must be explained by different astrophysics for it’s appearance in clusters that are not dynamically relaxed. Mass segregation is seen in clusters that are not dynamically relaxed, such as NGC 2071, the Trapezium, and NGC 2074 \citep{Lada91,Hillenbrand98,Bonnell98}. This mass segregation would be caused if higher mass stars tend to be formed nearer the center of clusters \citep{Proszkow09} or it could be an effect of dynamics during the formation of a not-yet dynamically relaxed star cluster as described by \citet{Maschberger10}. This phenomenon is also seen in W40. 
 
\begin{figure}
\centering
\includegraphics[angle=0.,width=7.0in]{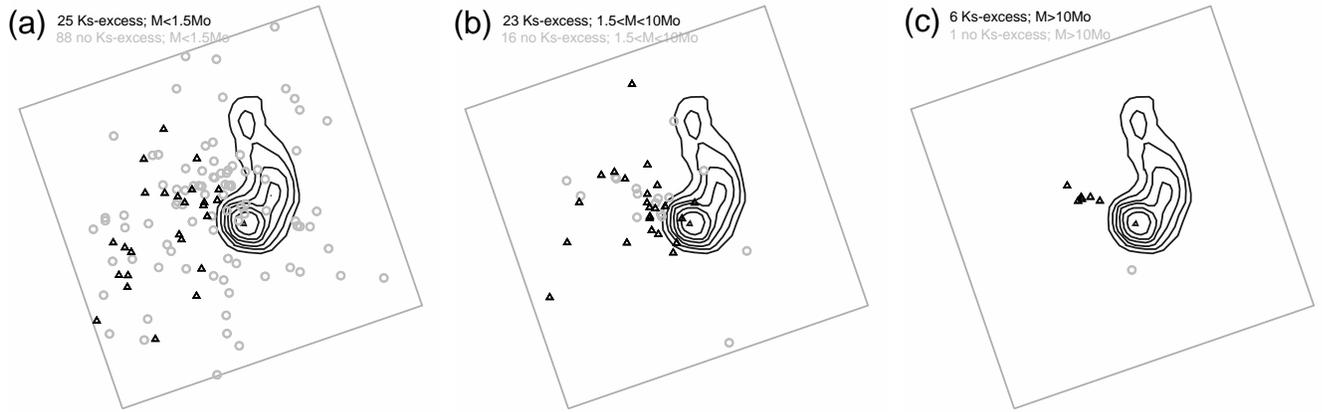}
\caption{Positions of W40 stars with (black triangles) and without (grey circles) $K_s$-excesses. The $17\arcmin \times 17\arcmin$ ACIS-I Field of View (grey box) is shown. The contour marks the $^{13}$CO emission from \citet{Zhu06}. Panels show different stellar mass strata: $M <1.5$~M$_{\odot}$ (a), $1.5<M<10$~M$_{\odot}$ (b), and $M>10$~M$_{\odot}$ (c). \label{fig_spat_distrib}}
\end{figure}

We stratify W40 cluster members by 3 mass ranges at 1.5~M$_{\odot}$ and 10~M$_{\odot}$ and by $K_s$-band excess (we will refer to the mass ranges as low mass, intermediate mass, and high mass). Both the intermediate-mass and high-mass populations should be approximately complete (based on the results of the XLF/IMF analysis). The spatial positions of stars in these strata are plotted in Figure \ref{fig_spat_distrib} and the centers and median radii are given in Table \ref{structure_table}. Low-mass stars are clearly less centrally concentrated than medium-mass stars and high-mass stars. The high-mass stars are the tightest group, except for star \#46 which is several arcminutes away from the rest of the group. We do not have proper motions to evaluate whether this star is a runaway. 

The CDFs of radial distances of low-mass stars, intermediate-mass stars, and high-mass stars show these mass segregation trends, and the K-S probabilities that the three mass distributions come from the same population are negligible\footnote{Each of the samples (low mass, intermediate mass, and high mass) is generated 1000 times using the same method described in \S \ref{imf_section} to take into account statistical and systematic errors on mass. The K-S test of the radial-distance CDFs is repeated for each of the 1000 simulations, and we find that more than 90\% of the significance levels are negligible for comparison of the low mass sample to the intermediate and high mass samples. The intermediate and high mass samples have marginally different CDFs in the simulated data.}. \citet{Moeckel09} describe a method that uses the area of the convex hull for $n$ stars to search for mass segregation. The convex hull is defined as the minimal convex polygon which covers a set of points, and the area may be used to determine stellar density. For W40, the probabilities that $n$ randomly selected low-mass stars could have a convex hull with an area less than the area of the convex hull of the $n$ high or intermediate-mass stars is $7\times10^{-4}$ and $5\times10^{-2}$, respectively. For the 25 stars with M$=$1.5$\pm$0.1~M$_{\odot}$ the probability is $2\times10^{-2}$.

This result is unusual since mass stratification typically is seen starting at higher masses than 1.5~M$_{\odot}$, for example \citet{Allison09} find no mass segregation in the ONC at masses less than 5~M$_{\odot}$. However, mass segregation beginning at 1~M$_{\odot}$ in 1~Myr old simulated clusters is found by \citet{Moeckel10} in an N-body simulation. 

The distance to W40 does not affect the existence of mass segregation, since radial distances scale similarly and the masses derived for 260~pc are an increasing function of the masses derived for 600~pc. However, mass segregation would starts at 0.3~M$_{\odot}$ for 260~pc rather than 1.5~M$_{\odot}$.

\subsection{Distribution of $K_s$-Excess Stars \label{section_ksdistribution}}

In Figure \ref{fig_spat_distrib}a the low mass stars with $K_s$-band excess are concentrated on the east side of the cluster, while the stars without $K_s$-band excess are evenly spread over the entire cluster. This effect is not as strong among the intermediate mass stars, many of which reside in the core from both populations. In Table \ref{structure_table} the median position for low-mass stars with $K_s$-band excess is 0.27$d$~pc to the east of the cluster center, which is farther away from the center than for any of the other populations listed. 

\begin{figure}
\centering
\includegraphics[angle=0.,width=7.0in]{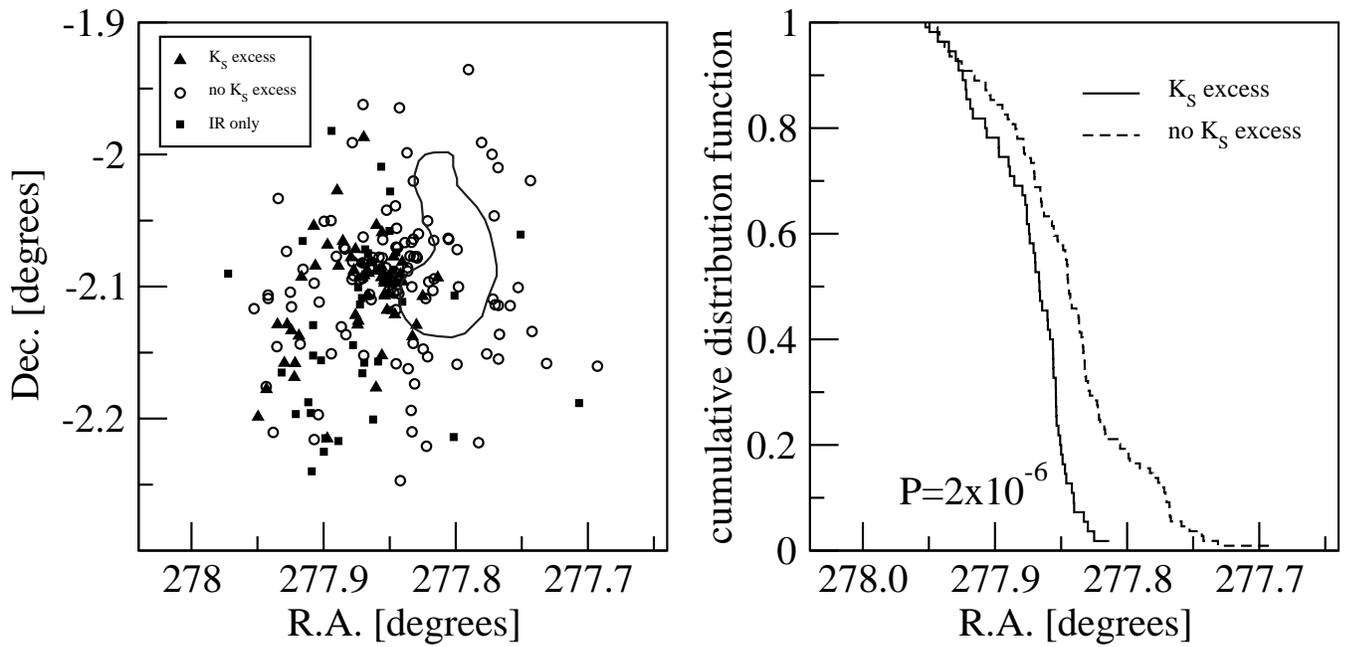}
\caption{(a) Positions of {\it Chandra}-detected $K_s$-excess stars (triangles) and non-$K_s$-excess stars (circles) and additional $K_s$-excess stars not detected in by {\it Chandra} (squares). The $^{13}$CO contour is also shown \citep{Zhu06}. (b) The cumulative distributions of {\it Chandra}-detected stars with and without $K_s$-excess are shown.\label{ksexcess_spatial_fig}}
\end{figure}

Figure \ref{ksexcess_spatial_fig} shows the spatial distribution of $K_s$-excess stars and the molecular core. Almost no stars with $K_s$-band excess are located west of the cluster core, and 20\% of the stars without $K_s$-band excess are located west of the westernmost $K_s$-excess star. The K-S probability that the spatial distributions are the same is $P=2\times10^{-6}$.


To the west of right ascension 18$^{\rm h}$ 31$^{\rm m}$ 15$^{\rm s}$ (277.81$^{\circ}$), 23 $Chandra$-detected stars have no $K_s$-band excess and 0 $Chandra$-detected stars have $K_s$-band excess, so the $1 \sigma$ upper limit on the observed $JHK_s$ disk fraction in this region is 5\% \citep[equation 21 in][]{Gehrels86}. East of 18$^{\rm h}$ 31$^{\rm m}$ 24$^{\rm s}$ (277.85$^{\circ}$), 45 $Chandra$-detected stars have no $K_s$-band excess and 35 $Chandra$-detected stars have $K_s$-band excess, so the observed $JHK_s$ disk fraction in this region is $\sim44^{+5}_{-7}$\%. Since only 20\% of T-Tauri stars with $K_s$-band excess are detected, compared to 45\% of T-Tauri stars without $K_s$-band excess, the astrophysical $JHK_s$ disk fraction in these two regions is $<10$\% and 63\%, for west and east regions respectively.

The extremely different disk fractions must have been caused recently, since this phenomenon will be erased due to mixing on the order of the cluster crossing time, which is $<1$~Myr. This provides further evidence that the age of the population of stars with $K_s$-band excess is $<$1~Myr.


Different $K_s$-band excess fractions may be attributed to environments hostile to disks or to different stellar ages. In the second case, our assumption of a coeval cluster is violated. This may be due to two different epochs of star formation or progressive star formation moving from west to east. Alternatively, the spatial distribution could be explained if the youngest stars moved to the east side of the cluster due to a similar initial velocity. Subclustering is also a possibility, although \S \ref{section_profile} gives evidence that the stars are from only one cluster.

A depleted disk fraction due to a harsh environment in the western part of the cluster seems unlikely, since the $K_s$-band excess disk fraction is high in the cluster core where the UV radiation and stellar winds would be strongest, and even some of the most massive stars have $K_s$-band excess. The disky stars in the core could have moved there recently, so they would not have been exposed to the environment for long enough to lose their disks. However, many of the disky stars in the core are high or intermediate mass stars, which are concentrated toward the center of the cluster (\S \ref{section_mass_segregation}).

\subsection{Dust Lane \label{section_highav}}

\begin{figure}
\centering
\includegraphics[angle=0.,width=6.5in]{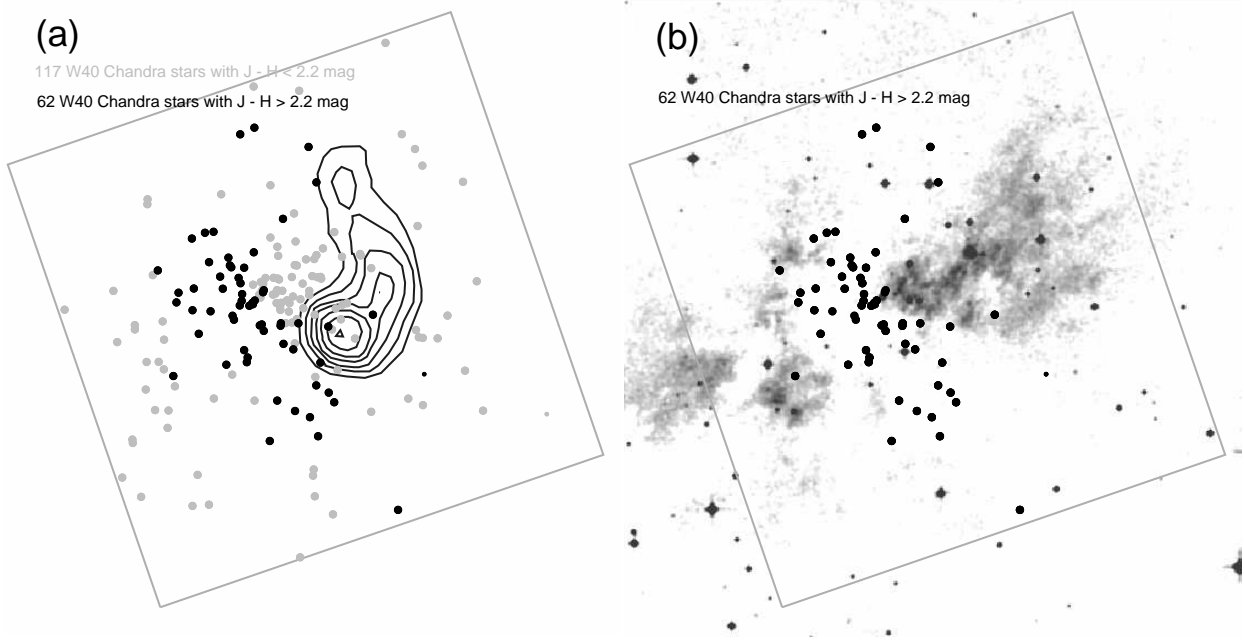}
\caption{(a) Positions of the W40 stars with $J-H>2.2$~mag (black circles) and $J-H<2.2$~mag (grey circles). The contour marks the $^{13}$CO emission from \citet{Zhu06}. (b) Positions of the W40 stars with $J-H>2.2$~mag over-plotted on the red Digitized Sky Survey image with inverted colors. The box outlines the {\it Chandra} ACIS-I field of view. \label{fig_high_absorption}}
\end{figure}

\begin{figure}
\centering
\includegraphics[angle=0.,width=6.5in]{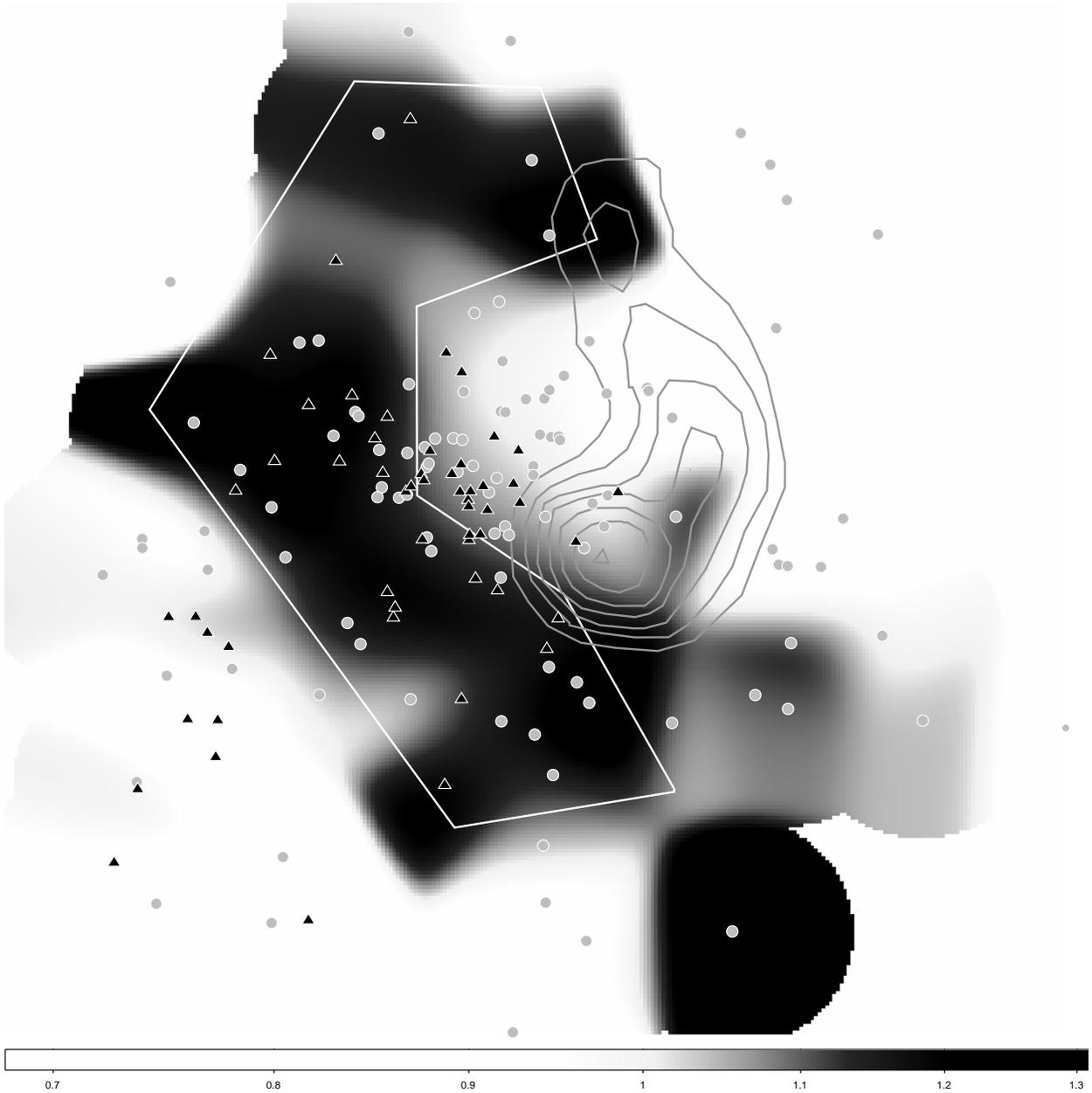}
\caption{The smoothed $A_V$ image is shown with intensity ranging from $\log(A_V)=$0.7 to 1.3~mag. Star positions are indicated by black triangles for stars with $K_s$-band excess and grey circles for stars without $K_s$-band excess. Contours show the molecular core in $^{13}$CO $J=2-1$ emission lines \citep{Zhu06}. A white polygon outlines the region with greatest $A_V$. 
\label{fig_av_map}}
\end{figure}

The MSX data show that W40 is a bipolar nebula with an hourglass shape \citep[Figure~6 in][]{Rodney08}. This type of structure is fairly common for H II regions; for example W40's shape closely resembles Sharpless 106 (S106) \citep{Oasa06}, IRAS 19343+2026 \citep{Ojha10}, and the Lagoon Nebula \citep{Tothill08}. In W40, the waist of the hourglass is the location of a dust lane which bifurcates the H II region and shows up in the optical image (Figure \ref{fig_high_absorption}b). Larger column densities of dust and gas are also measured for stars in the dark lane from the $JHK_s$ data and the X-ray data. The arrangement of the $J-H>$2.2 sources (corresponding roughly to $A_V \ga 10-13$~mag) is shown in Figure \ref{fig_high_absorption}a in comparison to the rest of the cluster and to the molecular core. We draw a region which encloses most of the $J-H>$2.2 sources, but few of the less absorbed sources, and plot this on a map of the smoothed extinction\footnote{The map of $A_V$ in the cluster is generated by smoothing the $A_V$ of stars with a Gaussian kernel with 1$^{\prime}$ standard deviation.} (shown in Figure \ref{fig_av_map}). The mean XPHOT hydrogen column density is $(3.2\pm0.4)\times10^{22}$~cm$^{-2}$ in the dark filament and $(1.8\pm0.1)\times10^{22}$~cm$^{-2}$ outside this region, considering only sources without $K_s$-band excess. These correspond to visual extinctions of 15~mag and 9~mag, respectively. The filament appears to have a normal gas-to-dust ratio: $\log N_H / \log A_V = 21.33\pm0.03$ for stars in the filament, and $\log N_H / \log A_V = 21.34\pm0.02$ for stars outside the filament.

The dust lane passes through the eastern half of the cluster core, and lies several arcminutes to the east of the CO core.  The filament has dimensions 3$^{\prime}$ by 12$^{\prime}$, and an approximate angular area of 36 arcmin$^2$ or 1.1$\times10^{37}$ cm$^2$ at a distance of 600~pc. The filament's mass is estimated by multiplying its angular area by the difference in mean column density inside and outside the lane giving a total mass of $\sim$200~M$_{\odot}$. This is comparable to the mass of the molecular core found by \citet{Zhu06}. If we assume the filament's extent along the line of sight is similar to its width of 3~arcmin, then the hydrogen density of the filament is $10^4$~atoms cm$^{-3}$. 

If the distance of 260~pc were considered instead, the masses of both the molecular core and dust lane would be scaled down by a factor of $\sim$5, but the density would be the same.

The presence of a large number of sources with $K_s$ excess near the dust lane, suggests that the lane may be a region of active star formation. This is further supported by the high concentration of {\it Herschel} protostar candidates from Figure 3 of \citet{Bontemps10} that lie in this region. In contrast, few deeply embedded X-ray sources are found in the CO core, suggesting that the molecular core is not the location of present star formation.

The geometry of the dust lane and the molecular core suggests that they form a ring, similar to the rings described by \citet{Beaumont10}. However, YSOs projected on the molecular core do not have very high $A_V$, and CO is not detected strongly in the dust lane. The former problem may be explained if the ring is oriented from our point of view so that the dust lane passes in front of the cluster and the molecular core is behind the cluster. The molecular core probably has a normal gas-to-dust ratio and we would see both CO emission and a visual dust lane if we were observing the cluster along a line of sight that passes through the cloud.

The lack of CO detection in the dust lane might mean that the molecular material has been destroyed in this region, or it might be an effect of the lack of sensitivity of the CO observations. CO may become frozen out onto dust grains in a molecular cloud \citep{vanDishoeck98}. \citet{Mitchell01} find evidence of this effect in two parts of the Orion B molecular cloud with bright dust continuum but almost absent CO emission. Alternatively, CO may be destroyed by UV radiation or expanding HII regions produced by OB stars. 

\section{Interesting X-ray Sources \label{interesting_section}}

\subsection{Stacked Point-Source Spectra \label{stacked_spectra}}

Out of 225 X-ray sources, only 24 sources have more than 100 net counts, and almost a third of the sources have fewer than 10 net counts. Composite spectra are useful for gaining insight into the sources which lack sufficient counts to be fit parametrically \citep[e.g.][]{Feigelson05}. They are created for the W40 point sources by merging individual source spectra, backgrounds, RMFs, and ARFs using the {\em ACIS Extract} software. Separate composite spectra are generated for cluster stars with more than 100 counts, cluster stars with fewer than 100 counts, and the extragalactic sources; these spectra are plotted in Figure \ref{fig_diffuse}b.

The spectra are fit in XSPEC using the thermal-plasma model $apec$ and the interstellar photoelectric-absorption model $wabs$. A two-temperature model with fixed parameters $Z=0.3$~Z$_{\odot}$ (see \S \ref{spectral_section}) and kT$_1$=0.8~keV (typical cool component in T-Tauri stars from \citet{Preibisch05b}) best fits the composite spectra of $>$100-count ($<$100-count) sources with $N_H=1.7\times10^{22}$~cm$^{-2}$ and kT$_2$=5.4~keV ($N_H=2.2\times10^{22}$~cm$^{-2}$ and kT$_2$=5.2~keV). For $>$100~count ($<$100~count) sources, model fitting with metal abundance $Z$ as a free parameter gives a value of $Z=$0.43$^{+0.09}_{-0.08}$ ($Z=$0.44$^{+0.14}_{-0.11}$), indicating that Z=0.3~Z$_{\odot}$ is a good approximation for the cluster. The stacked spectrum of extragalactic sources is well fit by an absorbed power law with a photon index of 1, within the typical range of $1-2$ for active galactic nuclei \citep{Alexander03}.

%
For the composite spectrum of $>$100-count ($<$100-count) sources XSPEC calculates observed (not corrected for absorption) hard-band fluxes of (2.9$\pm$0.1)$\times10^{-12}$~erg s$^{-1}$ cm$^{-2}$ ((1.7$\pm$0.1)$\times10^{-12}$~erg s$^{-1}$ cm$^{-2}$) and soft-band fluxes of (3.5$\pm$0.2)$\times10^{-13}$~erg s$^{-1}$ cm$^{-2}$ ((2.6$\pm$0.1)$\times10^{-13}$~erg s$^{-1}$ cm$^{-2}$), implying observed hard-to-soft flux ratios of 8.3$\pm$0.6 (6.5$\pm$0.5).

\subsection{The Diffuse X-ray Emission \label{diffuse_section}}

In a number of star forming regions the soft ($kT<1$~keV) diffuse thermal emission has been detected from a gas shocked in wind-wind or wind-surrounding medium interactions \citep{Townsley03,Gudel08}. The uncommon hard diffuse synchrotron emission has also been reported \citep[e.g.][]{Wolk02}. In some cases the observed X-ray diffuse emission can be completely explained by the unresolved stellar population \citep[e.g.][]{Getman06}.

\begin{figure}
\centering
\includegraphics[angle=0.,width=7.0in]{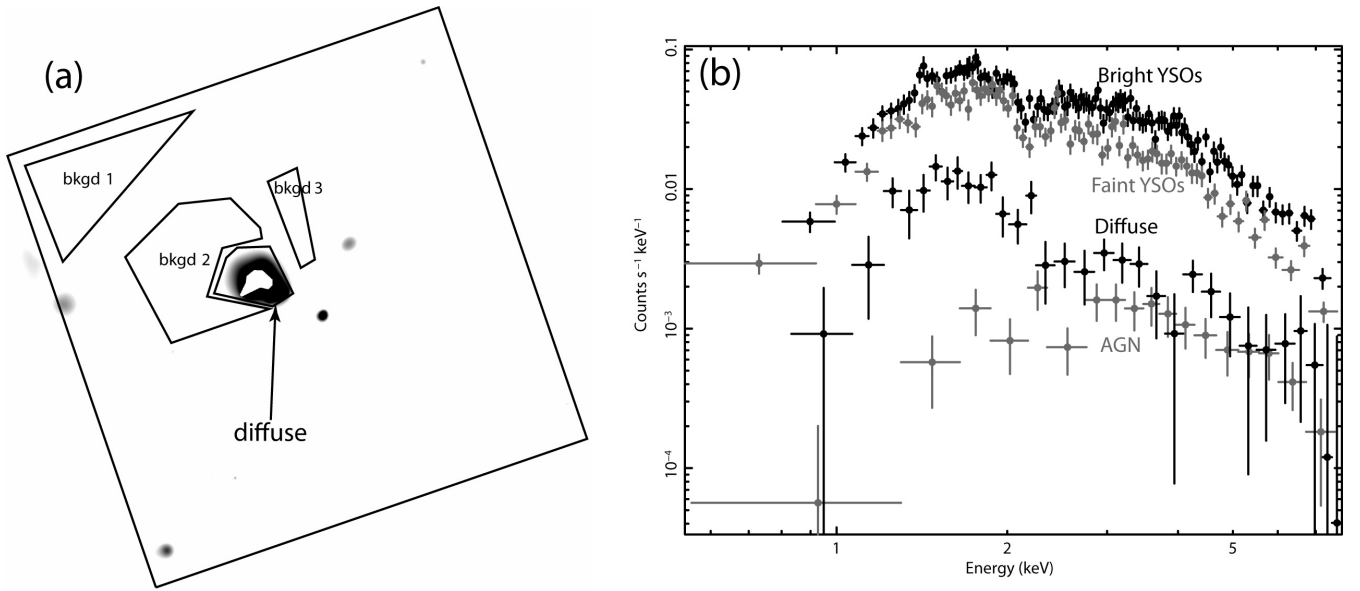}
\caption{(a) Total-band ($0.5-8$~keV) {\it Chandra} ACIS-I image after removal of point sources. This image is adaptively smoothed with a Gaussian kernel and normalized by the smoothed exposure map. The central polygon marks the extraction region for the diffuse X-ray emission with the carved out central part due to the contribution of the PSF wings of the point sources in the core of the cluster. Three background regions for the diffuse emission are also indicated. (b) Comparison of the composite spectra of the W40 stars, AGN, and the diffuse emission. \label{fig_diffuse}}
\end{figure}

The {\it Chandra} observation W40 has detected diffuse X-ray emission near the core of the cluster that is unassociated with the resolved point sources and can be seen in the smoothed image in Figure \ref{fig_cluster}. To isolate the diffuse emission, a ``Swiss-cheese'' mask is created \citep{Broos10}. Despite the Swiss-cheese masking, there is some light remaining from the wings of very bright X-ray sources concentrated in the core. To avoid any of this contamination a generously-sized polygon surrounding the central stars is excluded (Figure \ref{fig_diffuse}a). Three background regions are used; backgrounds 1 and 2 are on the the {\it Chandra} ACIS-I3 chip, while background 3 is on the {\it Chandra} ACIS-I2 chip. Background 2 is adjacent to the extraction region, while backgrounds 1 and 3 are further away\footnote{In the smoothed image with all point sources removed, there is a hint of extra diffuse emission to the south-west of the main diffuse emission. These excess counts lie near sources \#48 and \#51, within the molecular core. A possible origin could be a group of highly embedded PMS stars. However, spectral analysis suggests that the source spectrum is not significantly different from the local background.}.

%
%
%
%
A moderate-resolution spectrum is extracted with {\em ACIS Extract} from the source region and the three background regions. Fits with different backgrounds give similar results; reported results are given for background \#3. A degeneracy in spectral model fit exists due to the high absorption which makes it difficult to constrain the soft component ($kT<1$~keV) due to the small number of counts. A one-temperature model produces a good fit (reduced $\chi^2=$0.94) with $N_H=(1.2\pm0.2)\times10^{22}$~cm$^{-2}$, kT$=2.9\pm1.0$, and $Z=0.3$~Z$_{\odot}$. A variety of two-temperature models work as well, including $N_H=2.2\times10^{22}$~cm$^{-2}$, kT$_1=0.66$, kT$_2>15$, and $Z=0.3$~Z$_{\odot}$ and $N_H=3.3\times10^{22}$~cm$^{-2}$, kT$_1=0.27$, kT$_2=4.3$, and $Z=0.3$~Z$_{\odot}$, both with a reduced $\chi^2$ of $\sim0.8$. 

The observed (uncorrected for absorption) hard-band and soft-band fluxes of the diffuse emission are $(2.0\pm0.2)\times10^{-13}$~erg s$^{-1}$ and  $(4.8\pm0.1)\times10^{-14}$~erg s$^{-1}$, respectively, so the hard-to-soft flux ratio is $4.2\pm0.4$. This is smaller than that of the stacked stellar X-ray sources at a statistically significant level ($P<0.001$), which demonstrates that the spectra have a different shape. This can be seen visually in Figure \ref{fig_diffuse}b, where the bump at low energies is relatively larger in the diffuse spectrum than in the stacked stellar spectra. This result is an indication that the diffuse spectrum may not be entirely stellar. However, the unresolved stellar population, being composed of lower mass sources, is also expected to have spectra that are softer than the resolved population \citep[Figure 11 in][]{Preibisch05b}.

In the hard band only X-ray emission from stellar sources should contribute to the diffuse emission. The measured hard-band luminosity of the diffuse emission is compared with the expected hard-band luminosity of undetected stellar sources within the extraction region. 11\% of the {\it Chandra}-detected stars lie within the extraction region, so we assume a similar percentage of the unresolved stars lie in this region. The sum of the hard-band luminosity of all the missing sources from the XLF below the completeness limits is 6.8$\times10^{31}$~erg s$^{-1}$, and thus the hard-band emission expected within the extraction region is 7.5$\times10^{30}$~erg s$^{-1}$. The hard-band luminosity of the diffuse emission is calculated using the one-temperature and two-temperature models discussed above, and uncertainty is estimated by varying parameters within their 1~$\sigma$ errors. The result is $L_{hc}=(8.2\pm3.4)\times10^{30}$~erg~s$^{-1}$, which is consistent with that of the unresolved stellar component. 

The estimated total-band luminosity of undetected T-Tauri stars in the extraction region is $2.0\times10^{31}$~erg s$^{-1}$, which is calculated in the same way as the estimated hard-band luminosity. The total-band luminosity of the diffuse emission is $L_{tc}=(1.9\pm0.7)\times10^{31}$~erg~s$^{-1}$ for the one-temperature model fit, consistent with that of the undetected T-Tauri stars. However, it may be more than five times greater for the two-temperature model fits. Although, the total-band diffuse emission may be consistent with only the emission expected from unresolved stars, it is unclear if a real nonstellar X-ray component is present due to the degeneracy in spectral model fits.

If the distance to W40 is 260~pc, then diffuse emission luminosities decrease by a factor of 5. However, the IMF analysis indicates that the observation would be complete down to 0.1~M$_{\odot}$ so the X-ray emission from unresolved stars would be an order of magnitude lower in both the hard and the total bands.





\subsection{Super-Hot Flare in {\it Chandra} Star \#138 \label{section_flare}}

Long and powerful flares from YSOs are sometimes detected in {\it Chandra} observations of YSCs; for example, in the COUP project $\sim$200 flares are detected indicating a flare rate of approximately one per star per week \citep{Getman08}. During the 38~ks observation of W40, the decay phase of a powerful flare is detected in the light curve of source \#138, an intermediate mass (4~M$_{\odot}$) YSO near the center of the cluster. This flare has a peak luminosity $\ge$40 times its non-flaring luminosity and an e-folding decay timescale of 7.6~ks. The flare is strongly affected by pile-up on the ACIS-I chip, so we do not present flare cooling analysis. 

At least two alternative explanations for powerful flare activity from a 4~M$_{\odot}$ YSO exist. The object can be an F-type T-Tauri star with a partially developed radiative core. Big super-hot flares are typically seen from two types of T-Tauri stars: (1) low-mass stars ($M<2$~M$_{\odot}$) with active disks, (2) and more massive ($M>2$~M$_{\odot}$) disk-free stars \citep[\S 5 in][]{Getman08b}. It is possible that the formation of a radiative core in the latter case is associated with the change in the magnetic field dynamo, for example from turbulent to alpha-omega, which in turn strengthens their magnetic fields and allows production of super-hot flares. 

Alternatively, the object can be a fully radiative A or late B-type star. Although X-ray activity in late B and early A stars is generally weak, X-ray flares associated with these stars have been seen in several cases such as HD 161084 \citep{Miura08}, HD 261902, and HD 47777 \citep{Yanagida07}. 
One of the possibilities for the X-ray emission is the presence of an X-ray active T-Tauri companion to an IR-bright, A/B-type star.

\section{CONCLUSIONS \label{conclusions_section}}

We present a 38~ksec {\it Chandra} ACIS observation of the W40 complex. Although W40 is one of the nearest massive star-forming regions, it is poorly studied due to its high obscuration. Our main conclusions are as follows.

Two hundred twenty-five X-ray sources with a limiting luminosity of $\sim10^{29}$~erg s$^{-1}$ are detected (\S \ref{phot_section}), 202 of which are unambiguously identified with 2MASS or UKIRT NIR sources (\S \ref{identifications_section}). These along with 4 other sources (identified as T-Tauri stars by their variable X-ray emission) are classified as stellar sources. The 19 remaining sources are likely AGN (\S \ref{membership_section}). 

On the NIR color-magnitude diagram most of the detected X-ray stars occupy the locus of low-mass PMS stars. On the NIR color-color diagram 45 $K_s$-band excess and 120 non-$K_s$-band excess cluster members occupy the region with 5~mag$<A_V<$40~mag, but 10 stars are detected with low $A_V$ and are identified as foreground field stars (\S \ref{ir_properties_section}). The inferred X-ray properties are found to be consistent with IR properties of the stars through agreement with known $L_X$-mass and $N_H$-$A_V$ relations (\S \ref{confirmation_section}).

In the assumption of a universal XLF the XLF gives a lower limit distance to W40 of 600~pc (\S \ref{section_distance}). The XLF analysis shows that the total cluster population of PMS stars is $\ge$600 PMS stars assuming a distance of 600~pc, and a $K_s$-band excess disk fraction of 50\% suggests a cluster age $\le$1~Myr (\S \ref{tot_pop_section}). The IMF analysis is consistent with the XLF analysis on distance, total population, and disk fraction (\S \ref{imf_section}). The X-ray observation is not sensitive to a protostellar population that might exist in W40 \citep{Bontemps10}; only two protostellar X-ray candidates are identified (\S \ref{identifications_section}).

The young age of the W40 cluster, inferred from the high disk fraction and the existence of gas in the cluster, and the ongoing star formation in the dust lane (\S \ref{section_highav}) indicate that the cluster should not be dynamically relaxed yet (\S \ref{section_general_spatial}). However, the cluster does show some signs of relaxation, including a nearly spherical morphology with no elongation (\S \ref{section_elongation}), a surface density profile that may be described by an isothermal sphere (\S \ref{radial_profile}), and mass segregation (\S \ref{section_mass_segregation}). The cluster shape and profile may indicate the beginning of relaxation, but may also be geometric projection effects. Mass segregation down to 1.5~M$_{\odot}$ in W40 (\S \ref{section_mass_segregation}) is unusual, but it is predicted in simulation studies of YSCs by \citet{Moeckel10}.


The W40 star cluster may be composed of populations with different ages (\S \ref{section_ksdistribution}). Many of the stars without $K_s$-band excess may belong to an older population which has progressed further toward gravitational relaxation, while many of the stars with $K_s$-band excess, which are almost all located on the eastern side of the cluster, may be younger and may have not had time to mix with the rest of the cluster. It is likely that most of the high mass stars, 6 out of 8 having $K_s$-band excess, are members of the second population, suggesting delayed massive star formation (\S \ref{tot_pop_section}). 

The dust lane, which passes through the core and is detected by X-ray, optical, and IR absorption, may be the site of ongoing star formation (\S \ref{section_highav}). In contrast, no evidence is found for star formation in the molecular core detected in $^{12}$CO and $^{13}$CO. The dust lane and molecular core have similar masses and may both be part of a ring of interstellar material surrounding the star cluster. However, the dust lane is undetected in CO, and might be an example of the destruction of molecular gas by stars or molecular gas freezing out on to dust grains.

%
%
%
%
Diffuse X-ray emission is detected in W40 (\S \ref{diffuse_section}). However, it may be explained as X-rays from the unresolved stellar population. A powerful flare is observed from a $\sim$4~M$_{\odot}$ star without $K_s$-band excess (\S \ref{section_flare}). 


A minority of studies consider W40 to be part of the Aquila Rift at a distance of 260~pc, rather than the more commonly adopted distance of 600~pc (\S \ref{past_section}). The lower distance would modify results in the following ways: the shape of the W40 XLF would be inconsistent with the COUP XLF; for W40 stars without $K_s$-excess the shape of the IMF would be inconsistent with the \citep{Chabrier03} or COUP IMFs; the lower limit on the total cluster population would be 200 stars; only 2 stars would have masses above 10~M$_{\odot}$; and mass segregation would be seen down to 0.3~M$_{\odot}$. 

\acknowledgments We thank Patrick Broos and Leisa Townsley (Penn State) for their development of $Chandra$ tools and for helpful discussions about running {\em ACIS Extract}. We also thank them in addition to Kevin Luhman, Richard Wade, and Doug Cowan (Penn State) for giving advice on the paper. We thank the referee for many useful comments and suggestions. This work is supported by the $Chandra$ GO grant SAO G07-8010X (K. Getman, PI) and Chandra ACIS Team grant SV4-74018 (G. Garmire, PI). Additional support also comes from NASA grant NNX09AC74G and NSF grant AST-0908038. We use data products from the Two Micron All Sky Survey, a joint project of the University of Massachusetts and the Infrared Processing and Analysis Center/California Institute of Technology, which is funded by NASA and NSF. BR acknowledges support from the NASA Astrobiology Institute under Cooperative Agreement No. NNA08DA77A issued through the Office of Space Science.


\appendix
\section{Photometric Mass Uncertainty \label{mass_uncertainty_section}}

We make an attempt to evaluate systematic uncertainty on masses using a testbed sample of stars from the Taurus molecular cloud with known masses. The combined systematic and statistical error on masses of W40 stars is reported in Table \ref{tbl_irradio}. We do not attempt to estimate systematic uncertainty on stars with masses $>$5~M$_{\odot}$, and uncertainty on stars with masses up to 5~M$_{\odot}$ is assumed to be similar to that derived from the Taurus sample with masses between 0.5 and 1.5~M$_{\odot}$. Sources of error on stellar mass include statistical photometric error, uncertainty on distance, uncertainty on median cluster age, intrinsic age spread, reddening law and magnitude of extinction, model uncertainty including unmodeled effects of rotation, magnetic fields, and early accretion history \citep{Baraffe10,Chabrier07,Mohauty09,Jeffries08}, method uncertainty, binarity, and variability. 

A variety of evolutionary models are investigated by \citet{Hillenbrand04} who compare masses determined by placement on the Hertzsprung-Russell (H-R) diagram to dynamical masses for 27 PMS stars. For the \citet{Baraffe98} and \citet{Siess97} models used in our synthetic isochrones \citet{Hillenbrand04} found H-R masses are underestimated by $\sim$10\%. 

Ten out of the 27 PMS stars (7 single stars and 3 binaries) are from the Taurus star forming region with a mean distance of 140~pc and a median age of 2.5~Myr \citep[e.g.][]{Gudel07a}. The differences between dynamical mass and $JHK_s$ mass derived for these stars using the synthetic isochrone from 2.5~Myr and a distance of 140~pc, is used here to evaluate the magnitude of error on $JHK_s$ masses for W40. For binary stars we compare the $JHK_s$ mass to the mass of the most massive component. Table \ref{appendix_table} provides dynamical masses for these stars, 2MASS photometry, inferred $JHK_s$ masses, and the percent error on the $JHK_s$ masses.

The inferred difference in masses is around 30\%, adopted here as an average systematic error on mass for W40 stars. However, this is likely a lower limit on uncertainty. The uncertainties from intrinsic age spread, model uncertainty, method uncertainty, binarity, and variability are partially accounted for assuming similarity with the testbed of sample stars from Taurus. Distance, median age, and reddening law and extinction effects on mass uncertainty in W40 are not accounted for. Distance is fixed at 600~pc and 260~pc (masses using 260~pc are a factor of 3-4 lower then masses using 600~pc), age is fixed at 1~Myr (masses would be 60\% smaller using 0.5~Myr and a factor of 2 larger using 3~Myr), and the reddening law is assumed to be the \citet{Rieke85} law.

\clearpage

\begin{deluxetable}{cccccccccccccccc}
\centering \rotate \tabletypesize{\tiny} \tablewidth{0pt}
\tablecolumns{16}
\tablecaption{Basic Source Properties \label{tbl_phot}}
\tablehead{

\multicolumn{2}{c}{Source} &
\multicolumn{4}{c}{Position} &
\multicolumn{5}{c}{Extraction} &
\multicolumn{5}{c}{Characteristics} \\
                                
\multicolumn{2}{c}{\hrulefill} &  
\multicolumn{4}{c}{\hrulefill} &
\multicolumn{5}{c}{\hrulefill} &
\multicolumn{5}{c}{\hrulefill} \\

\colhead{No} & \colhead{CXOW40} &
\colhead{$\alpha$ (J2000.0)} & \colhead{$\delta$ (J2000.0)} & \colhead{Error} & \colhead{$\theta$} &
\colhead{$NC_{t}$} & \colhead{$\sigma NC_{t}$} & \colhead{$BG_{t}$} & \colhead{$NC_{h}$} & \colhead{PSF Frac.} &   
\colhead{Signif.} & \colhead{Anom.} & \colhead{Var.} &\colhead{Eff. Exp.} & \colhead{$ME$}  \\

\colhead{} & \colhead{} &
\colhead{(deg)} & \colhead{(deg)} & \colhead{(arcsec)} & \colhead{(arcmin)} &
\colhead{(counts)} & \colhead{(counts)} & \colhead{(counts)} & \colhead{(counts)} & \colhead{} &
\colhead{} & \colhead{} & \colhead{} & \colhead{(ks)} & \colhead{(keV)}
 \\

\colhead{(1)} & \colhead{(2)} &
\colhead{(3)} & \colhead{(4)} & \colhead{(5)} & \colhead{(6)} &
\colhead{(7)} & \colhead{(8)} & \colhead{(9)} & \colhead{(10)} & \colhead{(11)} &
\colhead{(12)} & \colhead{(13)} & \colhead{(14)} & \colhead{(15)} & \colhead{(16)}}

\startdata
160 & 183130.20-020718.0 & 277.875854 & -2.121672 & 0.1 &  2.1 &   17.8 &  4.8 &  0.2 &  17.9 & 0.91 &  3.3 & 0 & 2 & 34.5 & 4.3\\
161 & 183130.50-020515.8 & 277.877075 & -2.087735 & 0.1 &  1.8 &   29.7 &  6.0 &  0.3 &  23.8 & 0.90 &  4.5 & 0 & 4 & 34.6 & 3.2\\
162 & 183130.56-020530.6 & 277.877350 & -2.091831 & 0.1 &  1.7 &   10.8 &  3.8 &  0.2 &  10.9 & 0.90 &  2.4 & 0 & 2 & 33.0 & 3.2\\
163 & 183130.73-020452.2 & 277.878052 & -2.081161 & 0.2 &  2.0 &    4.9 &  2.8 &  0.1 &   0.9 & 0.70 &  1.4 & 0 & 2 & 34.6 & 1.8\\
164 & 183130.77-015927.8 & 277.878235 & -1.991042 & 0.4 &  6.7 &   36.5 &  6.8 &  2.5 &  28.4 & 0.89 &  5.0 & 0 & 4 & 27.1 & 2.8\\
165 & 183130.84-020540.6 & 277.878510 & -2.094621 & 0.2 &  1.8 &    5.8 &  3.0 &  0.2 &   4.9 & 0.90 &  1.6 & 0 & 3 & 33.3 & 2.8\\
166 & 183131.01-020440.5 & 277.879211 & -2.077922 & 0.1 &  2.2 &   30.7 &  6.1 &  0.3 &  24.8 & 0.89 &  4.6 & 0 & 2 & 34.5 & 2.8\\
167 & 183131.69-020611.6 & 277.882019 & -2.103231 & 0.2 &  2.0 &    8.8 &  3.5 &  0.2 &   4.8 & 0.91 &  2.1 & 0 & 2 & 34.4 & 2.2\\
168 & 183131.89-020011.6 & 277.882874 & -2.001029 & 0.6 &  6.2 &   10.7 &  4.1 &  2.3 &  10.5 & 0.89 &  2.3 & 0 & 2 & 31.0 & 3.7\\
169 & 183131.89-020611.6 & 277.882874 & -2.100822 & 0.2 &  2.0 &    5.8 &  3.0 &  0.2 &   4.8 & 0.91 &  1.6 & 0 & 2 & 34.4 & 3.7\\
\enddata

\tablecomments{Column 1: X-ray catalog sequence number, sorted by R.A. Column 2: IAU designation. Columns 3-4: Right ascension and declination for epoch (J2000.0). Column 5: Estimated standard deviation of the random component of the position error, $\sqrt{\sigma_x^2 + \sigma_y^2}$.  The single-axis position errors, $\sigma_x$ and $\sigma_y$, are estimated from the single-axis standard deviations of the PSF inside the extraction region and the number of counts extracted. Column 6: Off-axis angle. Columns 7-8: Net counts extracted in the total energy band (0.5--8~keV); average of the upper and lower $1\sigma$ errors on net counts. Column 9: Background counts expected in the source extraction region (total band). Column 10: Net counts extracted in the hard energy band (2--8~keV). Column 11: Fraction of the PSF (at 1.497 keV) enclosed within the extraction region. A reduced PSF fraction (significantly below 90\%) may indicate that the source is in a crowded region. Column 12: Photometric significance computed as net counts divided by the upper error on net counts. Column 13:  Source anomalies: ``1'' --- fractional time that source was on a detector (FRACEXPO from {\em mkarf}) is $<$0.9. Column 14: Variability characterization based on K-S statistic (total band): ``2'' --- no evidence for variability ($0.05<P_{KS}$); ``3'' --- possibly variable ($0.005<P_{KS}<0.05$); ``4'' --- definitely variable ($P_{KS}<0.005$).  Value ``1'' is reported for sources with fewer than four counts or for sources in chip gaps or on field edges. Column 15: Effective exposure time: approximate time the source would have to be observed at the aimpoint of the ACIS-I detector in Cycle 8 to obtain the reported number of net counts. Column 16: Background-corrected median photon energy (total band). Table \ref{tbl_phot} is published in its entirety in an electronic form. A portion is shown here for guidance regarding its form and content.}

\end{deluxetable}


\clearpage \clearpage

\begin{deluxetable}{ccccccccccccc}
\centering \rotate \tabletypesize{\tiny} \tablewidth{0pt}
\tablecolumns{13}
\tablecaption{IR and Radio Counterparts to {\it Chandra} Sources\label{tbl_irradio}}
\tablehead{

\colhead{No} & \colhead{CXOW40} & \colhead{2MASS} & \colhead{Offset}
& \colhead{$J$} & \colhead{$H$} & \colhead{$K_s$} & \colhead{PhCCFlg} & \colhead{MemFlg} &
\colhead{$A_V$} & \colhead{Mass} & \colhead{UKIRT\_Flg} & \colhead{VLA\_ID}\\

& & & (arcsec) & (mag) & (mag) & (mag) & & & (mag) & $M_{\odot}$ & &\\

(1)&(2)&(3)&(4)&(5)&(6)&(7)&(8)&(9)&(10)&(11)&(12)&(13)}

\startdata
160 & 183130.20-020718.0 & 18313020-0207180 &   0.0 & $ 16.61\pm$  0.15 & $ 13.59\pm$  0.04 & $ 11.66\pm$  0.03 & BAAcc0 & 	  W40-Ks	 &  20.09 &   1.44$\pm$0.53 & 1 &  20\\
161 & 183130.50-020515.8 & 18313049-0205158 &   0.0 & $ 15.70\pm$  0.05 & $ 13.05\pm$  0.04 & $ 11.62\pm$  0.03 & AAA000 & 	  W40-Ks	 &  17.25 &   1.48$\pm$0.47 & 1 & \nodata\\
162 & 183130.56-020530.6 & 18313056-0205306 &   0.1 & $ 15.57\pm$  0.07 & $ 13.09\pm$  0.02 & $ 11.82\pm$  0.02 & AAA00s & 	W40-noKs	 &  15.68 &   1.40$\pm$0.47 & 1 & \nodata\\
163 & 183130.73-020452.2 & 18313069-0204522 &   0.5 & $ 16.37\pm$  0.10 & $ 13.86\pm$  0.04 & $ 12.53\pm$  0.03 & BAA000 & 	W40-noKs	 &  14.09 &   0.58$\pm$0.19 & 1 & \nodata\\
164 & 183130.77-015927.8 & 18313077-0159275 &   0.2 & $ 16.72\pm$  0.15 & $ 14.34\pm$  0.04 & $ 13.24\pm$  0.04 & CAA000 & 	W40-noKs	 &  13.38 &   0.36$\pm$0.13 & 2 & \nodata\\
165 & 183130.84-020540.6 & 18313084-0205409 &   0.3 & $ 16.04\pm$  0.09 & $ 13.51\pm$  0.04 & $ 12.30\pm$  0.03 & AAA000 & 	W40-noKs	 &  14.72 &   0.82$\pm$0.29 & 1 & \nodata\\
166 & 183131.01-020440.5 & 18313101-0204405 &   0.1 & $ 14.55\pm$  0.03 & $ 11.57\pm$  0.02 & $  9.80\pm$  0.02 & AAA000 & 	  W40-Ks	 &  26.31 & 12.27$\pm$0.35 & 1 & \nodata\\
167 & 183131.69-020611.6 & 18313168-0206114 &   0.2 & $ 17.03\pm$\nodata & $ 14.20\pm$ 0.04 & $ 12.52\pm$  0.03 & UAA000 & 	     W40	 &  16.76 & \nodata & 1 & \nodata\\
168 & 183131.89-020011.6 & \nodata & \nodata & \nodata & \nodata & \nodata & \nodata &                                     	      EG	 & \nodata & \nodata & 0 & \nodata\\
169 & 183131.89-020611.6 & 18313187-0206022 &   0.7 & $ 18.37\pm$\nodata & $ 14.78\pm$ 0.05 & $ 13.44\pm$  0.04 & UAA000 & 	     W40	 &  25.37 & \nodata & 1 & \nodata\\
\enddata

\tablecomments{Columns 1-2: X-ray source number and IAU designation. Column 3: 2MASS source. Column 4: {\it Chandra}-2MASS positional offset. Columns 5-7: 2MASS $JHK_s$ magnitudes. Column 8: 2MASS photometry quality and confusion flag. Column 9: Membership flag: W40-noKS - W40 stars without $K_s$-excess; W40-KS - W40 stars with $K_s$-excess; FRG - foreground stars; EG - extragalactic source. Columns 10-11: Visual absorption and stellar mass derived from the color-magnitude $J$ vs. $J-H$ diagram assuming distance of $600$~pc and age of $1$~Myr. Column 12: UKIRT flag: 0 - no source; 1 - single source present; 2 - outside Field of View; 3 - double source present. Column 13: VLA source (Rodriguez, Rodney, \& Reipurth in preparation). Table \ref{tbl_irradio} is published in its entirety in an electronic form. A portion is shown here for guidance regarding its form and content.}

\end{deluxetable}

\clearpage \clearpage

\begin{deluxetable}{ccccccc}
\centering \rotate \tabletypesize{\tiny} \tablewidth{0pt}
\tablecolumns{7}
\tablecaption{Additional Non X-ray Cluster Members \label{tbl_additional}}
\tablehead{

\colhead{2MASS} &  \colhead{$J$} & \colhead{$H$} & \colhead{$K_s$} & \colhead{PhCCFlg} &
\colhead{$A_V$} & \colhead{Mass} \\

& (mag) & (mag) & (mag) & & (mag) & $M_{\odot}$\\

(1)&(2)&(3)&(4)&(5)&(6)&(7)}

\startdata

 18304958-0211176 & $ 16.07\pm$0.11 & $ 14.62\pm$0.06 & $ 13.67\pm$0.05 & BAA000 &        6.43 &   0.12	$\pm$0.08\\
 18310021-0203383 & $ 16.50\pm$0.15 & $ 14.16\pm$0.04 & $ 12.53\pm$0.03 & CAA000 &       12.67 &   0.38	$\pm$0.04\\
 18311221-0206246 & $ 16.46\pm$0.13 & $ 14.69\pm$0.06 & $ 13.54\pm$0.05 & CAA000 &        8.94 &   0.15	$\pm$0.05\\
 18311237-0212503 & $ 16.48\pm$0.14 & $ 15.27\pm$0.11 & $ 14.34\pm$0.09 & CBA0cc &        4.62 &   0.07	$\pm$0.14\\
 18312171-0206416 & $ 16.53\pm$0.18 & $ 14.53\pm$0.10 & $ 12.82\pm$0.08 & CAA000 &       10.30 &   0.19	$\pm$0.05\\
 18312271-0205484 & $ 15.85\pm$0.09 & $ 14.42\pm$0.05 & $ 13.49\pm$0.05 & AAAccc &        6.11 &   0.13	$\pm$0.06\\
 18312276-0204269 & $ 16.01\pm$0.09 & $ 14.14\pm$0.02 & $ 12.97\pm$0.02 & AAAc00 &        9.28 &   0.23	$\pm$0.04\\
 18312330-0205139 & $ 16.32\pm$0.11 & $ 14.04\pm$0.07 & $ 12.45\pm$0.03 & BAAccc &       12.06 &   0.38	$\pm$0.39\\
 18312391-0201407 & $ 15.89\pm$0.08 & $ 13.49\pm$0.03 & $ 12.00\pm$0.02 & AAA000 &       13.23 &   0.68	$\pm$0.04\\
 18312396-0207040 & $ 16.84\pm$0.18 & $ 14.93\pm$0.07 & $ 13.66\pm$0.08 & CAAccc &       10.16 &   0.15	$\pm$0.04\\
18312403-0203282 & $ 16.07\pm$  0.08 & $ 14.27\pm$  0.04 & $ 13.05\pm$  0.04 & AAA000 &   8.79 &   0.19	$\pm$0.34\\
18312554-0200332 & $ 17.01\pm$  0.18 & $ 15.43\pm$  0.10 & $ 14.32\pm$  0.08 & CAA000 &   7.69 &   0.09	$\pm$0.05\\
18312611-0209246 & $ 15.63\pm$  0.05 & $ 13.46\pm$  0.03 & $ 12.16\pm$  0.03 & AAA000 &  11.19 &   0.54	$\pm$0.22\\
18312701-0212024 & $ 16.51\pm$  0.13 & $ 14.06\pm$  0.04 & $ 12.53\pm$  0.03 & BAA000 &  13.49 &   0.46	$\pm$0.08\\
18312701-0204530 & $ 16.45\pm$  0.12 & $ 14.74\pm$  0.07 & $ 13.46\pm$  0.05 & BAAccc &   8.38 &   0.14	$\pm$0.04\\
18312793-0204290 & $ 16.14\pm$  0.10 & $ 14.42\pm$  0.05 & $ 13.36\pm$  0.04 & AAAccc &   8.38 &   0.16	$\pm$0.03\\
18312846-0204187 & $ 16.55\pm$  0.13 & $ 14.90\pm$  0.06 & $ 13.67\pm$  0.06 & BAA000 &   7.95 &   0.12	$\pm$0.05\\
18312862-0209275 & $ 15.68\pm$  0.07 & $ 14.18\pm$  0.03 & $ 13.22\pm$  0.04 & AAA000 &   6.57 &   0.16	$\pm$0.03\\
18312894-0209565 & $ 14.80\pm$  0.03 & $ 12.80\pm$  0.02 & $ 11.61\pm$  0.02 & AAA000 &   9.93 &   0.74	$\pm$0.06\\
18312902-0206314 & $ 16.94\pm$  0.20 & $ 14.98\pm$  0.06 & $ 13.70\pm$  0.07 & CAA000 &  10.52 &   0.15	$\pm$0.16\\
18312919-0204551 & $ 15.31\pm$  0.04 & $ 12.97\pm$  0.02 & $ 11.56\pm$  0.04 & AAAccc &  13.42 &   1.05	$\pm$0.03\\
18312938-0206490 & $ 15.36\pm$  0.05 & $ 13.17\pm$  0.04 & $ 11.86\pm$  0.03 & AAA000 &  11.37 &   0.68	$\pm$0.17\\
18312972-0206022 & $ 16.00\pm$  0.09 & $ 13.38\pm$  0.04 & $ 11.66\pm$  0.03 & AAA000 &  15.62 &   1.03	$\pm$0.05\\
18313065-0208398 & $ 16.45\pm$  0.12 & $ 14.92\pm$  0.06 & $ 13.76\pm$  0.04 & BAAccc &   7.00 &   0.11	$\pm$0.06\\
18313331-0213012 & $ 16.18\pm$  0.10 & $ 14.34\pm$  0.04 & $ 13.02\pm$  0.03 & AAA000 &   9.11 &   0.19	$\pm$0.02\\
18313457-0158555 & $ 15.90\pm$  0.08 & $ 13.91\pm$  0.04 & $ 12.72\pm$  0.03 & AAA000 &  10.00 &   0.32	$\pm$0.04\\
18313566-0212534 & $ 16.52\pm$  0.11 & $ 15.11\pm$  0.08 & $ 14.09\pm$  0.07 & BAAccc &   6.29 &   0.09	$\pm$0.02\\
18313595-0213305 & $ 16.11\pm$  0.09 & $ 13.97\pm$  0.04 & $ 12.32\pm$  0.02 & AAA000 &  10.90 &   0.34	$\pm$0.11\\
18313646-0209212 & $ 14.37\pm$  0.04 & $ 12.30\pm$  0.04 & $ 10.97\pm$  0.02 & AAA000 &  12.09 &   1.40	$\pm$0.03\\
18313789-0209080 & $ 16.75\pm$  0.13 & $ 15.39\pm$  0.10 & $ 14.35\pm$  0.12 & CABccc &   5.83 &   0.08	$\pm$0.03\\
18313791-0207453 & $ 16.34\pm$  0.10 & $ 14.44\pm$  0.04 & $ 13.23\pm$  0.03 & BAA000 &   9.75 &   0.19	$\pm$0.23\\
18313815-0214236 & $ 16.80\pm$  0.14 & $ 14.77\pm$  0.06 & $ 13.48\pm$  0.05 & CAA000 &  10.91 &   0.17	$\pm$0.06\\
18313832-0211449 & $ 16.05\pm$  0.07 & $ 14.70\pm$  0.04 & $ 13.82\pm$  0.04 & AAA000 &   5.62 &   0.11	$\pm$0.03\\
18313878-0211155 & $ 16.45\pm$  0.10 & $ 14.85\pm$  0.06 & $ 13.76\pm$  0.06 & BAA0c0 &   7.63 &   0.12	$\pm$0.03\\
18313979-0203558 & $ 16.99\pm$  0.17 & $ 15.29\pm$  0.10 & $ 14.10\pm$  0.07 & CAA000 &   8.53 &   0.10	$\pm$0.43\\
18314106-0211475 & $ 16.73\pm$  0.15 & $ 15.65\pm$  0.12 & $ 14.72\pm$  0.12 & CBB000 &   3.58 &   0.06	$\pm$0.14\\
18314362-0209545 & $ 16.54\pm$  0.10 & $ 15.38\pm$  0.08 & $ 14.28\pm$  0.07 & BAA000 &   4.15 &   0.07	$\pm$0.23\\
18315335-0205248 & $ 17.02\pm$  0.19 & $ 15.43\pm$  0.08 & $ 14.34\pm$  0.08 & CAA0cc &   7.77 &   0.09	$\pm$0.10\\
OS~3a            & $  8.91\pm$  0.03 & $  8.02\pm$  0.05 & $  7.37\pm$  0.02 & AAA0dd &   8.84 &   15.00  $\pm$0.72\\	
\enddata

\tablecomments{Column 1: 2MASS source. Columns 2-4: 2MASS $JHK_s$ magnitudes. Column 5: 2MASS photometry quality and confusion flag. Columns 6-7: Visual absorption and stellar mass derived from the color-magnitude $J$ vs. $J-H$ diagram assuming distance of $600$~pc and age of $1$~Myr.}

\end{deluxetable}

\clearpage \clearpage

\begin{deluxetable}{cccccccccccccc}
\centering \rotate \tabletypesize{\tiny} \tablewidth{0pt}
\tablecolumns{14}
\tablecaption{X-ray Spectroscopy \label{tbl_spectroscopy}}
\tablehead{

\multicolumn{5}{c}{Source} &
\multicolumn{6}{c}{XSPEC} &
\multicolumn{3}{c}{XPHOT} \\
                                
\multicolumn{5}{c}{\hrulefill} &  
\multicolumn{6}{c}{\hrulefill} &
\multicolumn{3}{c}{\hrulefill} \\

\colhead{No} & \colhead{CXOW40} & \colhead{$NC$} & \colhead{$ME$}
& \colhead{MemFlg} & \colhead{$\log(L_{h1})$} & \colhead{$\log(L_{t1})$}& \colhead{$kT_1$} & \colhead{$\log(N_{H1})$} &
\colhead{$\log(L_{hc1})$} & \colhead{$\log(L_{tc1})$} & \colhead{$\log(N_{H2})$} & \colhead{$\log(L_{hc2})$} & \colhead{$\log(L_{tc2})$}\\

& & (cnts) & (keV) & & (ergs~s$^{-1}$) & (ergs~s$^{-1}$) & (keV) & (cm$^{-2}$) & (ergs~s$^{-1}$) & (ergs~s$^{-1}$) & (cm$^{-2}$) & (ergs~s$^{-1}$) & (ergs~s$^{-1}$)\\

(1)&(2)&(3)&(4)&(5)&(6)&(7)&(8)&(9)&(10)&(11)&(12)&(13)&(14)}

\startdata
160 & 183130.20-020718.0 &   17.85 & 4.30 & 	  W40-Ks	 & 29.79 & 29.79 &  1.73 & 23.26 & 30.52 & 31.02 & $23.15\pm 0.19$ & $30.31\pm 0.17$ & $30.73\pm 0.24$\\
161 & 183130.50-020515.8 &   29.74 & 3.22 & 	  W40-Ks	 & 29.86 & 29.89 &  5.03 & 22.46 & 29.99 & 30.22 & $22.64\pm 0.19$ & $30.07\pm 0.14$ & $30.52\pm 0.27$\\
162 & 183130.56-020530.6 &   10.77 & 3.23 & 	W40-noKs	 & 29.47 & 29.47 &  1.53 & 22.96 & 29.99 & 30.57 & $22.71\pm 0.26$ & $29.74\pm 0.20$ & $30.22\pm 0.36$\\
163 & 183130.73-020452.2 &    4.88 & 1.76 & 	W40-noKs	 & 28.62 & 28.86 &  0.55 & 22.54 & 28.99 & 30.65 & \nodata & \nodata & \nodata\\
164 & 183130.77-015927.8 &   36.54 & 2.84 & 	W40-noKs	 & 29.88 & 29.92 &  1.23 & 22.71 & 30.27 & 31.00 & $22.50\pm 0.15$ & $30.09\pm 0.12$ & $30.56\pm 0.25$\\
165 & 183130.84-020540.6 &    5.82 & 2.76 & 	W40-noKs	 & 29.09 & 29.12 &  1.72 & 22.66 & 29.38 & 29.88 & $22.52\pm 0.28$ & $29.25\pm 0.23$ & $29.76\pm 0.34$\\
166 & 183131.01-020440.5 &   30.73 & 2.83 & 	  W40-Ks	 & 29.75 & 29.79 &  1.76 & 22.56 & 30.00 & 30.50 & $22.50\pm 0.14$ & $29.99\pm 0.12$ & $30.41\pm 0.24$\\
167 & 183131.69-020611.6 &    8.78 & 2.17 & 	     W40	 & 29.11 & 29.18 &  1.83 & 22.37 & 29.27 & 29.74 & $22.30\pm 0.26$ & $29.23\pm 0.21$ & $29.72\pm 0.29$\\
168 & 183131.89-020011.6 &   10.67 & 3.72 & 	      EG	 & 29.48 & 29.48 &  1.39 & 22.97 & 30.04 & 30.67 & $22.88\pm 0.28$ & $29.88\pm 0.21$ & $30.33\pm 0.32$\\
169 & 183131.89-020611.6 &    5.78 & 3.66 & 	     W40	 & 29.13 & 29.15 &  1.83 & 22.75 & 29.46 & 29.93 & $22.89\pm 0.34$ & $29.57\pm 0.28$ & $30.10\pm 0.40$\\
\enddata

\tablecomments{Columns 1-5: For convenience, these columns reproduce the source identification, net counts, median energy, and membership flag from Tables \ref{tbl_phot} and \ref{tbl_irradio}. Columns 6-11: Apparent X-ray luminosities, plasma temperature, column density, and intrinsic X-ray luminosities inferred from XSPEC fitting. Columns 12-14: X-ray column density, intrinsic luminosities, and their errors (summed in quadrature statistical and systematic errors) obtained with XPHOT.pro. X-ray luminosities are derived assuming a distance of 600~pc. Table \ref{tbl_spectroscopy} is published in its entirety in an electronic form. A portion is shown here for guidance regarding its form and content.}

\end{deluxetable}

\clearpage \clearpage

\begin{deluxetable}{ccccccc}
\centering 
\tabletypesize{\tiny} \tablewidth{0pt}
\tablecolumns{7}
\tablecaption{Completeness Limits and Scale Factors \label{table_xlf_completeness}}
\tablehead{

&
\multicolumn{2}{c}{Hard-Band XLF} &
\multicolumn{2}{c}{Total-Band XLF} &
\multicolumn{2}{c}{IMF} \\

\colhead{Subpopulation} & \colhead{log Completeness Limit} & \colhead{Scaling} & \colhead{log Completeness Limit} & \colhead{Scaling}& \colhead{Completeness Limit} & \colhead{Scaling}\\

& (erg s$^{-1}$) & & (erg s$^{-1}$) & & (M$_{\odot}$) & \\

(1)&(2)&(3)&(4)&(5)&(6)&(7)}

\startdata
all  w40           &30.2        &         0.70&     30.6       &    0.70  & 1.5 & 0.72 \\
no $K_s$-band excess &\nodata    &           0.30&     30.0-30.4  &    0.35& 0.5    &          0.32\\
$K_s$-band excess    &\nodata        &         0.30&     30.8       &    0.35& 1.5            &      0.40\\

\enddata

\tablecomments{Column 1: Population examined. We examine the entire sample of W40 cluster members, cluster members with $K_s$-band excess, and cluster members without $K_s$-band excess. Columns 2,4,6: The completeness limits at which the scaled COUP XLF model and \citet{Chabrier03} model depart from the W40 XLF and IMF, respectively. Columns 3,5,7: The scale between the model needed to fit the COUP sample of $>800$ stars and the model needed to fit the W40 subpopulation.}

\end{deluxetable}

\clearpage \clearpage

\begin{deluxetable}{ccccccccc}
\centering 
\tabletypesize{\tiny} \tablewidth{0pt}
\tablecolumns{9}
\tablecaption{NIR and X-ray Properties of Early-Type Star Candidates \label{OB_table}}
\tablehead{

\colhead{No} & \colhead{Name} & \colhead{$NC$} & \colhead{Mass} & \colhead{$\log(L_{\rm Bol})$} & \colhead{$K_s$-band excess} & \colhead{$N_H$} & \colhead{kT} & \colhead{$\log(L_{tc})$}\\

& & (cnts) & (M$_{\odot}$) & (L$_{\odot}$) & & (10$^{22}$~cm$^{-2}$) & (keV) & (ergs~s$^{-1}$)\\

(1)&(2)&(3)&(4)&{5}&(6)&(7)&(8)&(9)}

\startdata
      46  &   183117.12-020911.5 &        9     &    12   &     4.11   &   NoKs     &        2.32  & 1.64  & 30.06 \\
      99  &   183124.00-020529.5, OS 2a & 58    &    30   &     5.25   &   Ks     &        3.99  & 1.24  & 30.95\\
     122  &   183126.02-020517.0 &        1177  &    16   &     4.46   &   Ks     &        1.69  & 3.55  & 31.63\\
     141  &   183127.84-020523.5, OS 1a & 145   & $<$32   &  $<$5.36   &   Ks     &        1.78  & 1.21  & 30.94\\
     144  &   183128.01-020517.1 &        11    &    14   &     4.24   &   Ks    &        1.19  & 0.68  & 29.84\\
     145  &   183128.67-020529.8 &        417   &    21   &     4.85   &   Ks     &        3.41  & 3.70  & 31.42\\
     166  &   183131.01-020440.5 &        41    &    12   &     4.09   &   Ks    &        9.38  & 0.90  & 31.47\\
  \nodata &   OS 3a  &                \nodata   &    15   &     4.36   &   NoKs &      \nodata & \nodata & \nodata \\
\enddata

\tablecomments{Columns 1-2: X-ray source number and IAU designation. Column 3: X-ray source net counts. Columns 4-5: NIR photometric stellar mass and bolometric luminosity. Column 6: $K_s$-band excess indicator. Columnns 7-9: Inferred X-ray spectral properties.}

\end{deluxetable}

\clearpage \clearpage

\begin{deluxetable}{cccc}
\centering 
\tabletypesize{\tiny} \tablewidth{0pt}
\tablecolumns{4}
\tablecaption{Cluster Structure \label{structure_table}}
\tablehead{

\colhead{Subpopulation} & \colhead{Center Distance} & \colhead{Center Angle} & \colhead{Median Radius}\\

& (pc) & (degrees) & (pc)\\

(1)&(2)&(3)&(4)}

\startdata
ALL      &  0.00 & NA  &0.50\\
LM       &  0.02 & 319 &0.60\\
IM       &  0.01 & 270 &0.36\\
HM       &  0.12 & 107 &0.08\\
NKE      &  0.11 & 273 &0.59\\
KE       &  0.12 & 80  &0.33\\
LM; NKE  &  0.13 & 276 &0.66\\
IM; NKE  &  0.08 & 232 &0.41\\
LM; KE   &  0.27 & 62  &0.54\\
IM; KE   &  0.04 & 348 &0.35\\
\enddata

\tablecomments{Column 1: The subpopulation of stars used for structure analysis; LM - low mass stars ($M<1.5$~M$_{\odot}$); IM - intermediate mass stars ($1.5<M<10$~M$_{\odot}$); HM - high mass stars ($M>10$~M$_{\odot}$); NKE - stars with no $K_s$-band excess; KE - stars with $K_s$-band excess. Column 2: Distance from the cluster center to the subpopulation center (median position). Column 3: Position angle of the subpopulation center with respect to the cluster center. Column 4: The radius of the circle around the subpopulation center that contains half the stars in the subpopulation.}

\end{deluxetable}


\clearpage \clearpage
\clearpage

\begin{deluxetable}{ccccccc}
\centering 
\tabletypesize{\tiny} \tablewidth{0pt}
\tablecolumns{7}
\tablecaption{Dynamical and $JHK_s$ Masses of Taurus Stars \label{appendix_table}}
\tablehead{

\colhead{Name} & \colhead{$J$} & \colhead{$H$} & \colhead{$K_s$} & \colhead{$M_{{\rm Dyn}}$} & \colhead{$M_{JHK_s}$} & \colhead{\% Error}\\

& (mag) & (mag) & (mag) & (M$_{\odot}$) & (M$_{\odot}$) & \\

(1)&(2)&(3)&(4)&(5)&(6)&(7)}

\startdata

MWC 480 &  5.41 &  5.34 &  5.34 & 1.65$^a$ & 2.68 & 62\\
BP Tau &  9.10 &  8.22 &  7.74 & 1.32$^b$ & 1.22 &  -7\\
LkCa15 &  9.42 &  8.60 &  8.16 & 0.97$^a$ & 0.73 &  -24\\
GM Aur &  9.34 &  8.60 &  8.28 & 0.84$^a$ & 0.66 &  -21\\
DL Tau &  9.63 &  8.68 &  7.96 & 0.72$^a$ & 0.78 &  8\\
DM Tau & 10.44 &  9.76 &  9.52 & 0.55$^a$ & 0.30 &  -44\\
CY Tau &  9.83 &  8.97 &  8.60 & 0.55$^a$ & 0.58 &  6\\
045251+30168A &  8.96 &  8.32 &  8.13 & 1.45$^c$ & 0.98 &  -32\\
045251+30168B &  8.96 &  8.32 &  8.13 & 0.81$^c$ & \nodata &  \nodata \\
UZ Tau Aa &  8.45 &  7.60 &  7.35 & 1.02$^a$ & 2.10 & 105\\
UZ Tau Ab &  8.45 &  7.60 &  7.35 & 0.294$^a$ & \nodata & \nodata \\
GG Tau Aa &  8.67 &  7.82 &  7.36 & 0.67$^{ad}$ & 1.87 & 180\\
GG Tau Ab &  8.67 &  7.82 &  7.36 & 0.61$^{ad}$ & \nodata & \nodata \\

\enddata

\tablecomments{Column 1: Star name. Columns 2-4: 2MASS $JHK_s$ magnitudes. Column 5: Dynamical masses. Column 6: 2MASS photometric masses. Column 7: Percent error in photometric masses. $^a$ \citet{Simon00}. $^b$ \citet{Dutrey03}. $^c$ \citet{Steffen01}. $^d$ \citet{White99}.
}

\end{deluxetable}
\clearpage \clearpage


\clearpage

\clearpage

\clearpage

\clearpage


\clearpage

\clearpage



\clearpage

\clearpage

\clearpage

\clearpage

\clearpage

\clearpage


\clearpage

\clearpage

\clearpage

\clearpage


\end{document}